\documentclass[
 reprint,
superscriptaddress,
 amsmath,
 amssymb,
 aps,
twocolumn,
longbibliography
]{revtex4-2}
\usepackage{dcolumn}
\usepackage[utf8]{inputenc}
\usepackage{graphicx,bm,color}
\usepackage[dvipsnames]{xcolor}
\usepackage{amsthm,amsmath,amssymb,mathtools}
\usepackage{multirow}
\usepackage{colortbl}
\usepackage{wrapfig}
\usepackage{braket}
\definecolor{britishracinggreen}{rgb}{0.0, 0.5, 0.15}
\definecolor{burgundy}{rgb}{0.5, 0.0, 0.13}
\definecolor{egyptianblue}{rgb}{0.06, 0.2, 0.65}
\usepackage[colorlinks=true, citecolor=burgundy, linkcolor=britishracinggreen, urlcolor=egyptianblue]{hyperref}
\usepackage{enumitem}
\usepackage{ragged2e}
\usepackage{float}

\newcommand{\hncdi}{The Hartree Centre, STFC, Sci-Tech Daresbury, Warrington WA4 4AD, UK}

\newcommand{\ibmde}{IBM Quantum, IBM Research Europe -- Ehningen, Germany}

\newcommand{\ibm}{IBM Quantum, IBM Research Europe -- Zurich, Switzerland}

\usepackage{fbm_usra_macros}
\graphicspath{{./figures/}}

\begin{document}
\nocite{apsrev42Control}
\title{Quantum Approximate Optimization\\ via Noise-Directed Adaptive Warm-Starting}
\author{Filip B. Maciejewski}
\email{fmaciejewski@usra.edu}
\author{Stuart Hadfield}
\affiliation{\riacs}
\author{Oscar Wallis}
\author{George Pennington}
\affiliation{\hncdi}
\author{Sebastian Brandhofer}
\affiliation{\ibmde}
\author{Stefan Woerner}
\author{Daniel J. Egger}
\affiliation{\ibm}
\author{Davide Venturelli}
\affiliation{\riacs}

\date{\today}

\begin{abstract}
    Progress towards a quantum advantage using known heuristic methods for combinatorial optimization is impeded by hardware noise and limited qubit count. Here, we propose a noise-aware adaptive approach to quantum approximate optimization, Noise-Directed Adaptive Warm-Starting (ND-AWS), that builds on recent concepts such as Warm-Start QAOA and Noise-Directed Adaptive Remapping. By leveraging bitflip gauge transformations, our algorithm exploits amplitude-damping-like noise components. We experimentally implement high-performance quantum optimization ansätze on 100-qubit Ising Hamiltonians, showing that ND-AWS generally improves the performance over a non-gauge-transformed iterative Warm-Starting variant, at no additional circuit cost. This places our results among the highest-quality demonstrations of quantum optimization with similar ansätze at this scale. Crucially, the simplicity of the framework opens the door for future enhancements such as adaptive bias schedules, and integration with classical solvers.
\end{abstract}

\maketitle

\section{Introduction}

Quantum computing hardware has progressed to the point where we can test quantum algorithms at a scale beyond the reach of exact classical simulation methods. A promising area where quantum computing may deliver value is combinatorial optimization~\cite{abbas2023quantum}. Despite increases in qubit count, gate quality, and hardware speed, the quantum circuits that can be meaningfully executed are still limited in depth and width by noise and qubit count. As a result, quantum optimization algorithms such as the Quantum Approximate Optimization Algorithm (QAOA)~\cite{farhi2014quantum,hadfield2019quantum} and its variants are typically restricted to noisy, low-depth instantiations on current quantum hardware, with corresponding limitations on performance. This is an issue since many algorithms for optimization---both quantum and classical---are heuristic and therefore require careful testing and benchmarking on hardware~\cite{koch2025decathlon}.
Hence, a number of recent works seek a more sophisticated use of limited quantum resources. 
For example, recursive (a.k.a. iterative) quantum 
optimization 
algorithms utilize local expectation values to progressively simplify the target problem through variable reduction~\cite{bravyi2020obstacles,dupont2023quantum,brady2023iterative,finvzgar2024quantum, brady2025quantum}. 
Closely related are several dual
approaches~\cite{Dupont2024QRR,dupont2025optimization,vcepaite2025quantum} that propose to improve the performance of classical solvers by directly leveraging quantum results.
Warm-Start variants of QAOA directly bias the quantum circuit using approximate or relaxed problem solutions, and can achieve better results with fewer resources than standard QAOA~\cite{egger2021warm, tate2023bridging, tate2023warm, cain2022qaoa, augustino2024strategies, okada2024systematic, chai2024structure, yuan2024quantifying, dehn2024hybrid, 
grass2019quantum, yu2022quantum, yu2023solution,yu2025warm,feeney2025better,bhattacharyya2025solving,ha2025difference}, most recently in an iterative manner~\cite{yuan2025iterative,lopez2025non,marshall2026quantum, bucher2026constrainedquantumoptimizationiterative}. 
The same ideas apply to other ansätze~\cite{lotshaw2026IWSIT} that can be
warm started and are not limited to QAOA and its variants.

Our proposed approach seeks to exploit both algorithmic bias, as well as that induced from hardware noise.
Many quantum algorithms can be executed in ideal hardware in multiple equivalent ways, due to logical symmetries of the formal circuit mapping. For example, in typical combinatorial optimization  approaches, the ``1'' value of a binary variable $x\in\{0, 1\}$ of the cost function can equivalently be mapped to either of the two levels offered by a qubit implementation (and vice versa for the ``0''). 
However, noise in quantum hardware often breaks this symmetry, like when the two levels correspond to the ground or excited state of a qubit, which are differently susceptible to decoherence and dissipation effects.
Noise-Directed Adaptive Remapping (NDAR)~\cite{maciejewski2024ndar,maciejewski2024multilevel,tam2025enhancingNDAR} leverages this bitflip gauge by gauge transforming the cost and phase separation operators in QAOA circuits to align the optimization with the ideal (zero-temperature) ground state of the quantum device. 
This allows the algorithm to exploit amplitude-damping-like noise.

Here, we synergize iterative warm-starting and NDAR into a new noise-adaptive approach to quantum approximate optimization we denote Noise-Directed Adaptive Warm-Starting (ND-AWS). 
We use an iterative procedure, where repeated preparation and measurement yield a distribution of solutions. 
At each iteration, the Warm-Start ansatz is updated to bias it towards the current best solution.
Here, we align the best solution with the hardware noise via bitflip gauge transformations, making the algorithm inherently adapted to amplitude-damping-like noise.
The procedure continues with a bias strength schedule until a termination criterion is met (such as a high-quality solution has been obtained).

\begin{figure*}[!t]
\centering
\includegraphics[width=\textwidth]{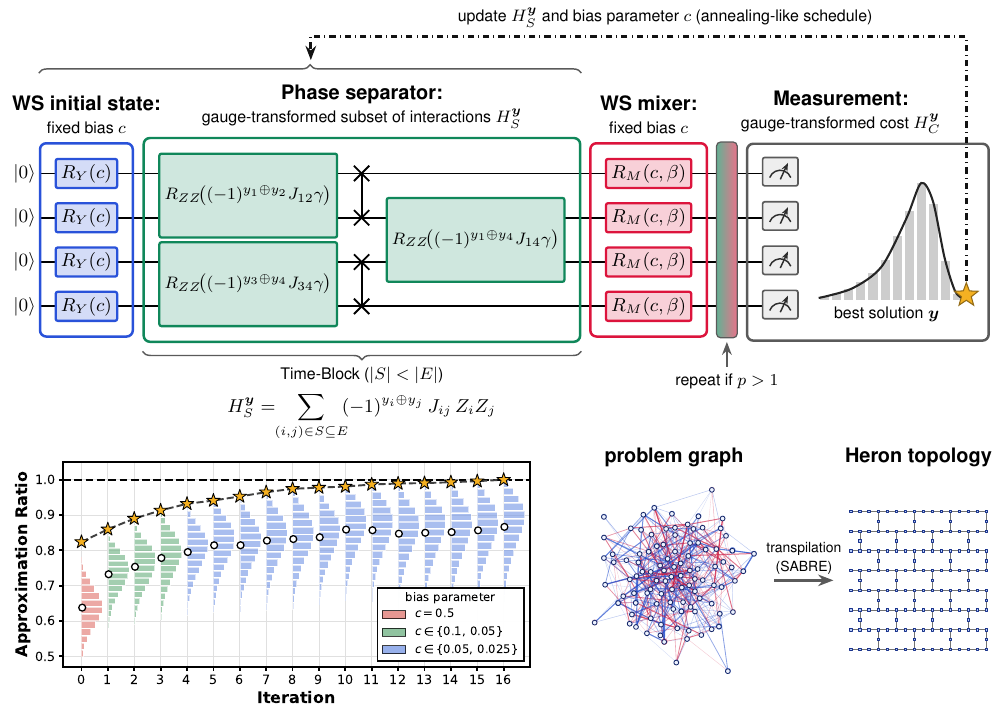}
\caption{\label{fig:illustration_IWS}
Illustration of the Noise-Directed Adaptive Warm-Starting algorithm implementation introduced in this paper.
\textbf{(top)} In the vanilla Warm-Start QAOA~\cite{egger2021warm}, the phase separator Hamiltonian is equal to the cost Hamiltonian, and fixed during the execution.
Our shallow-depth ansatz expands upon Warm-Start QAOA by combining it with Time-Block QAOA~\cite{maciejewski2023design}, and Noise-Directed Adaptive Remapping~\cite{maciejewski2024ndar}.
The Time-Block ansatz divides the original cost Hamiltonian $H_C = \sum_{S}H_{S}$ into batches of interactions $H_{S}$, meaning only some interactions from the cost Hamiltonian are implemented in each layer.
The initial state (Eq.~\eqref{eq:ws_ininital_state}) and mixer Hamiltonian (Eq.~\eqref{eq:ws_mixer_one_qubit}) are fixed at each iteration, and always biased towards the ideal thermal ground state of the Quantum Processing Unit (QPU), $\ket{0\dots0}$.
The degree of bias is controlled by the $c$ parameter, the same for all qubits.
The gradient-colored block indicates that the phase separator and mixer are re-applied (with possibly different parameters) for depth $p>1$.
At the end of each iteration, the best-found solution $\mathbf{y}$ (star) is fed back (dash-dotted arrow) to update both the warm-start bias and the gauge of the cost and Phase Separator (PS) Hamiltonians.
\textbf{(bottom left)} Sampling from the QPU at each iteration (x-axis) results in distributions of solutions of varying quality (y-axis); distributions are binned per iteration and colored according to the bias-parameter schedule.
In the ND-AWS loop, we gauge-transform the cost and PS Hamiltonians at each iteration based on the results of the previous iteration.
This causes the next-iteration distribution to be moved towards the best-found solution from the previous iteration, focusing the search in the higher-quality regions of solution space.
\textbf{(bottom right)} An exemplary $100$-qubit problem graph (edge color and intensity illustrate the sign and magnitude of the couplings $J_{ij}$) is routed to the heavy-hexagonal topology of the QPU with the SABRE algorithm~\cite{li2019sabre,zou2024lightsabre}.
}
\end{figure*}

We test the ND-AWS experimentally by sampling from $100$ qubits of \texttt{ibm\_boston}, a superconducting quantum processing unit (QPU).
We consider $20$ random Hamiltonian instances with a connectivity given by Erd\H{o}s-R\'enyi graphs with $10\%$ (ER-10) and $20\%$ (ER-20) edge probability; as well as $10$ instances of random $3$-regular (RG-3) graphs.
The resulting approximation ratios, obtained in a variant with additional greedy classical post-processing at each iteration, lie in the $0.974$--$1.0$ range for ER-10, $0.97$--$1.0$ for ER-20, and $0.989$--$1.0$ for RG-3, when the best out of $3$ independent QPU runs is considered.
In particular, we test against an
iterative Warm-Starting variant similar to that of Ref.~\cite{lopez2025non}
that does not employ gauge transformations, and conclude that the bitflip transforms in ND-AWS generally improve the optimization quality.
To the best of our knowledge, our results are among the highest quality results to date for combinatorial problems solved by QAOA variants with around 100 qubits
~\cite{abbas2023quantum,Pelofske2024ScalingQAOA,wang2025linearchainQAOA, mohseni2026copulaQAOA, montanezbarrera2025evaluatingperformancequantumprocessing, montanezbarrera2025LinearRampQAOA, rava2025benchmarkingneutralatombasedquantum}.
Moreover, Matrix Product State (MPS) approximate simulations indicate that better performance will be achievable once higher circuit depths become experimentally feasible.

The rest of this paper is structured as follows.
Section~\ref{sec:iws} presents our algorithm and the specific realizations we implement. 
We elaborate on the connections to NDAR and related methods in Section~\ref{sec:related_work}.
A numerical proof of concept of the noise adaptivity of our approach is shown in Section~\ref{sec:numerics}, with further supporting simulations provided in the appendices.
Section~\ref{sec:exp} presents our 100-qubit experimental implementation on \texttt{ibm\_boston}.
Finally, we discuss our results and future work in Section~\ref{sec:discussion}.

\section{Noise-Directed Adaptive Warm-Starting}\label{sec:iws}

\subsection{General Algorithm}

We seek candidate solutions (classical bitstrings) $x$ that minimize the energy $\langle H_C\rangle=\langle x| H_C|x\rangle$ of a cost Hamiltonian $H_C$.
Our algorithm samples $x$ from a gauge-transformed ansatz circuit biased towards the zero-temperature ground state of the device, $\ket{0\dots 0}$.
Thanks to the gauge transformation, that state becomes logically equivalent to the best solution of the previous iteration.
At each iteration, we
\begin{enumerate}
\item optimize variational parameters of the ansatz (e.g., offline in simulations),
    \item sample on quantum hardware from the ansatz,
  \item gauge-transform $H_C$ so that the (ideal) ground state of the QPU is logically equivalent to the best-found sample (see Eq.~\eqref{eq:gauge_transformation_local}),
  \item optionally, update the hyperparameters of the method, such as the WS bias value $c$ (see Eqs.~\eqref{eq:ws_ininital_state} and \eqref{eq:ws_mixer_one_qubit}),
  \item move to the next iteration.
\end{enumerate}
The procedure stops when it meets a termination criterion, for example, a shot budget is reached or the best solution does not improve over the past few iterations~\cite{neira2024benchmarking}. 
Fig.~\ref{fig:illustration_IWS} illustrates our algorithm.
As the utilized quantum resources increase, we expect our approach to explore increasingly non-local neighborhoods of the previous best sample, providing a mechanism for potential quantum advantage.  

At each iteration, our quantum optimization effectively starts in the neighborhood of the previous best-found solution, similar to many classical local-search algorithms.
At the very first iteration, we run standard QAOA without bias since we do not have any knowledge of what a good solution might be.
Naturally, one could also start the feedback loop with a bias towards a classically found solution. 

\subsection{Quantum ansätze}\label{sec:ansatz}

Consider the \emph{minimization} of a classical Hamiltonian
\begin{align}\label{eq:cost_hamiltonianZZ}
    H_C = \sum_{\left(i,j\right) \in E} J_{ij} Z_iZ_j 
\end{align}
where $Z_i$ is the Pauli $Z$ operator acting on qubit $i$, and $J_{i j} \in \mathbb{R}$ is the interaction strength between qubits $i,j$ in the edge set $E$. 
This Hamiltonian captures problems such as MaxCut and the Sherrington-Kirkpatrick (SK) model of spin glasses~\cite{sherrington1975solvable}.
Throughout the paper, when discussing random Hamiltonian instances, the $J_{ij}$ will be taken from the normal distribution with mean $0$ and $\sigma=1$. 
For simplicity, we focus on the cost Hamiltonian of Eq.~(\ref{eq:cost_hamiltonianZZ}), but our algorithm also applies to any optimization problem encoded as a linear combination of Pauli $Z$ operators~\cite{lucas2014ising,hadfield2021representation}, see~App.~\ref{app:warm_starting_qaoa}.

To find low-energy states of $H_C$, we iteratively sample candidate bitstrings from the ansatz circuit
\begin{align}\label{eq:ansatz}
    \ket{\psi_{\y}} = \left(\prod_{l=p}^{1}U\left(H_M;\beta_l\right)\ U_{\mathrm{PS}}\left(H^{\y}_{\mathrm{PS},l};\gamma_l\right)\right)  \ket{\mathbf{c}} \ .
\end{align}
Here, $U(H;\alpha)=\exp(-i\alpha H)$, and $\gamma_{l},\beta_{l} \in \mathbb{R}$, and in presented simulations and experiments we use $p=1$.

Typically, the phase separator $H_\textrm{PS}$ in the ansatz is the same as the cost Hamiltonian $H_C$.
However, due to noise, currently implementing even a single layer of such a phase separator is often infeasible.
Therefore, we implement a \emph{Time-Block (TB)} ansatz~\cite{maciejewski2023design} with phase separator 
\begin{align}
    H_{\mathrm{PS},l} = \sum_{(i,j) \in \mathcal{S}_{l}}
    J_{ij} Z_iZ_j .
\end{align}
Here, the $\mathcal{S}_{l} \subseteq E$ are subsets of the edges $E$ selected distinctly for each step $l$ such that their union after some fixed number of steps gives~$E$, see Appendix~\ref{app:time_block_ansatz}.
Importantly, in practice, we often do not reach enough layers to cover the whole graph.
Compared to the standard QAOA, where we have $\mathcal{S}_{l}=E$ and $l=1$, this choice reduces the circuit expressivity with the gain of reducing physical depth, and thus the impact of noise. 

Importantly, we align the phase separator $H^{\y}_{\mathrm{PS}}$ and initial state $\ket{c}$ with the ideal ground state of the QPU such that the algorithm and common noise sources act in the same direction. 
At each iteration $r$ except the first one, the cost and phase Hamiltonians $H_C$ and $H_{\mathrm{PS}}$ are thus gauge-transformed by a candidate solution $\y$ from the previous iteration $r-1$.
The gauge transformation works as follows~\cite{maciejewski2024ndar}. 
Consider a candidate solution $\ket{\y} = \ket{y_0,\dots,y_{\noq-1}}$; with $y_i \in \left\{0,1\right\}$, and a corresponding unitary bitflip operator $P_{\y} = \bigotimes_{i=0}^{\noq-1}X_{i}^{y_i}$ that flips the $|0\rangle$ and $|1\rangle$ basis states when bit $y_i$ is $1$.
The corresponding change-of-basis is applied to $H_{\mathrm{PS}}$ and $H_{C}$ 
by changing the weights $J_{ij}$ since 
\begin{align}\label{eq:gauge_transformation_local}
 H^{\y} 
    = P_{\y}HP_{\y}
= \sum_{i,j} \left(-1\right)^{y_i\oplus y_j}J_{ij} Z_iZ_j \ .
\end{align}
This transformation preserves the Hamiltonian eigenvalues, with eigenvectors (candidate problem solutions) permuted under $P_{\y}$. 
Importantly, for QAOA circuits the transformation is straightforward to implement in hardware. 
As seen from Eq.~\eqref{eq:gauge_transformation_local}, it requires only changing signs in the right gates of $U_\text{PS}$. 
The cost Hamiltonian measurement basis change is done in post-processing.
Under this transformation the $|0\dots0\rangle$ state is mapped to~$|y_0\dots y_{n-1}\rangle$ such that
\begin{align}
    \bra{\y}H\ket{\y}=\bra{0\dots 0}H^{\y}\ket{0\dots 0}\ .
\end{align}
Therefore, the bitstring $\y$  from iteration $r-1$ maps to the QPU's zero-temperature ground state $\ket{0\dots0}$ at iteration $r$.
To explore candidate solutions in the neighborhood of $\y$, we warm-start iteration $r$ with a state biased towards $\ket{0\dots0}$, and, thanks to the gauge-transformations, that state is logically equivalent to the best candidate solution at iteration $r-1$.
We call this choice the \emph{Noise-Directed (ND)} ansatz.
In general, any classical state could be used in place of $\ket{0\dots 0}$ -- we use the ND ansatz due to the expected adaptivity towards amplitude-damping-like noise, following Ref.~\cite{maciejewski2024ndar}.

Each qubit is initialized in the state
\begin{align}\label{eq:ws_ininital_state}
\ket{c} \coloneqq R_{Y}\left(\theta_c\right)\ket{0} = \sqrt{1-c}\ket{0} + \sqrt{c} \ket{1}
\end{align}
where $\theta_c=2\arcsin(\sqrt{c}) \in \left[0, \pi\right]$, and the full state on $\noq$ qubits is $\ket{c}^{\otimes \noq}$.
Here, the \emph{bias parameter} $c \in \left[0.0, 0.5\right]$ controls the degree of biasing towards the $\ket{0}$ state. Standard QAOA corresponds to $c=0.5$. 
Since the mixer operator should have the initial state $\ket{c}$ as ground state, see, e.g., Ref.~\cite{he2023alignment}, we use $H_{M} = \sum_{i} H_{i}^{M}$
with
\begin{align}\label{eq:ws_mixer_one_qubit}
    H_{i}^{M} = \left(2\sqrt{c\left(1-c\right)}\right)X_i+\left(1-2c\right)Z_i \ .
\end{align} 
Here, we use the angle-sign convention such that $\ket{c}$ is the ground state of $-H_M$.
Without the gauge transformation, we would implement a qubit-dependent initial state $\ket{c_i}$ and mixer $H_{i}^{M}\left(c_i\right)$, where the value of $c_i$ for each qubit depends on the solution bit $y_i$ while keeping $H_{C}$ and $H_{\mathrm{PS}}$ fixed~\cite{egger2021warm,lopez2025non}.
In contrast, our algorithm moves the bias from the best solution onto $H_C$ and $H_{\mathrm{PS}}$.
When comparing against the ND variant, we will sometimes refer to this non-gauge-transformed choice as \emph{Standard} Warm-Starting.
In a noiseless setting, both ansätze result in the same probability distributions upon measurement, see Appendix~\ref{app:qaoa_symmetries}.

Finally, the parameters $\left\{\gamma_l\right\}_{l=0}^{p-1}$  and $\left\{\beta_l\right\}_{l=0}^{p-1}$  are chosen to minimize
\begin{align}
\left\langle \psi_\y(\boldsymbol{\beta}, \boldsymbol{\gamma}) | H_C^\y | \psi_\y(\boldsymbol{\beta}, \boldsymbol{\gamma}) \right\rangle.
\end{align}
Here, $H_C^\y$ is the gauge-transformed cost operator.

\subsection{Performance metrics}

Our performance metric is the approximation ratio
\begin{equation}\label{eq:approximation_ratio}
    AR = \frac{E_{\max}-E}{E_{\max}-E_{\min}}\, .
\end{equation}
Here, $E$ is the energy of the Hamiltonian $H$ that we wish to \emph{minimize}.
We find the lowest and the highest energies $E_{\min}$ and $E_{\max}$ with classical solvers.
The \emph{worst} and \emph{best} candidate solutions have AR $0.0$ and $1.0$ and correspond to the highest and lowest energy states, respectively.
For visualization purposes, when solutions are close to optimal, we will use $1-AR$.

Unless stated otherwise, we compute the $AR$ with respect to solutions found by either the Burer-Monteiro (BM) algorithm~\cite{burer2002BM_original} implemented in \texttt{mqlib}~\cite{dunning2018mqlib} and \texttt{quapopt}~\cite{quapopt_repo} or a custom-implemented TABU search solver~\cite{TABUprivate,quapopt_repo}.
Burer-Monteiro is a heuristic rank-2 relaxation of the Goemans-Williamson SDP MaxCut solver~\cite{goemans1995GW_original}. 

\subsection{Related approaches}\label{sec:related_work}
Iteratively improving a given problem solution is a common theme in classical heuristics.
Recent work extends this idea to the quantum domain. 
These approaches similarly use samples obtained from the quantum device to iteratively update the quantum ansatz.
In particular, Ref.~\cite{yuan2024quantifying} considers iteratively warm-started initial states that are superpositions of best-found solutions from previous iterations; while in Ref.~\cite{lopez2025non}, the bias parameters of the WS-ansatz of~\cite{egger2021warm} are iteratively updated in the Iterative-QAOA algorithm. 
The approach of Refs.~\cite{yuan2024quantifying,lopez2025non} was recently extended in Refs.~\cite {marshall2026quantum, bucher2026constrainedquantumoptimizationiterative}.
For simplicity of exposition, we refer to similar methods as Iterative Warm-Starting (IWS). 
Ref.~\cite{lotshaw2026IWSIT} alternatively proposes an iterative scheme using a quantum circuit ansatz related to imaginary-time evolution.
Distinctly, our work additionally builds on NDAR~\cite{maciejewski2024ndar,maciejewski2024multilevel, tam2025enhancingNDAR}, which uses the current best solution found to iteratively update the problem encoding (i.e., gauge) as well as the quantum ansatz and bias.

Beyond the Warm-Start QAOA of~\cite{egger2021warm}, there are a number of suitable warm start variants~\cite{tate2023bridging, tate2023warm, he2023alignment, cain2022qaoa, augustino2024strategies, okada2024systematic, chai2024structure, yuan2024quantifying, dehn2024hybrid, yu2025warm,feeney2025better,bhattacharyya2025solving,ha2025difference,carmo2025warmXY, carmo2025warmTSP, truger2024warm, truger2024warmstarting} which could be incorporated into our approach. 
In the context of quantum annealing, the paradigm of reverse annealing~\cite{developers2021reverse,ohkuwa2018reverse,kechedzhi2018efficient,marshall2019power,venturelli2019reverse,mehta2025unraveling} is closely related to circuit model Warm-Start approaches and may be amenable to iterative schemes.
Further afield, though similar in spirit to our approach, several works propose algorithms involving adaptively chosen bias parameters. 
QAOA is modified to include local fields with iteratively updated biases in \cite{yu2022quantum,yu2023solution}, and combined with warm starting in \cite{yu2025warm}, with a related approach for quantum annealing previously proposed in \cite{grass2019quantum}.
Finally, we note that \cite{barton2025iterative} gives a further related approach, where sampled bitstrings are iteratively used to set spin-dependent phases for the oscillating drive terms therein.

\begin{table}[h!]
\centering
\resizebox{\columnwidth}{!}{%
\begin{tabular}{cccc}
\hline
\multicolumn{1}{|c|}{\textbf{Cost Hamiltonian}}          & \multicolumn{1}{c|}{ER-10}                        & \multicolumn{1}{c|}{ER-20}                        & \multicolumn{1}{c|}{RG-3}                         \\ \hline
                                                         &                                                   &                                                   &                                                   \\ \hline
\multicolumn{1}{|c|}{\textbf{Number Of Qubits}}          & \multicolumn{3}{c|}{100}                                                                                                                                  \\ \hline
\multicolumn{1}{|c|}{\textbf{Number Of Edges}}           & \multicolumn{1}{c|}{$499\ \left(\pm 33\right)$}   & \multicolumn{1}{c|}{$993\ \left(\pm 40\right)$}   & \multicolumn{1}{c|}{$150$}                        \\ \hline
\multicolumn{1}{|c|}{\textbf{Average Degree}}            & \multicolumn{1}{c|}{$10\ \left(\pm 0.7\right)$}   & \multicolumn{1}{c|}{$19.9\ \left(\pm 0.8\right)$} & \multicolumn{1}{c|}{$3$}                          \\ \hline
\multicolumn{1}{|c|}{\textbf{Phase Separator$^\ast$}}    & \multicolumn{1}{c|}{$25\%$}                       & \multicolumn{1}{c|}{$10\%$}                       & \multicolumn{1}{c|}{$80\%$}                       \\ \hline
\multicolumn{1}{|c|}{\textbf{Algorithmic depth}}         & \multicolumn{3}{c|}{$p=1$}                                                                                                                                \\ \hline
                                                         &                                                   &                                                   &                                                   \\ \hline
  \multicolumn{1}{|c|}{\textbf{$\sqrt{X}$ Gate Count}}     & \multicolumn{1}{c|}{$2769\ \left(\pm 285\right)$} & \multicolumn{1}{c|}{$1935\ \left(\pm
  147\right)$} & \multicolumn{1}{c|}{$2394\ \left(\pm 104\right)$} \\ \hline
  \multicolumn{1}{|c|}{\textbf{$CZ$ Gate Count}}           & \multicolumn{1}{c|}{$1146\ \left(\pm 140\right)$} & \multicolumn{1}{c|}{$736\ \left(\pm
  72\right)$}   & \multicolumn{1}{c|}{$963\ \left(\pm 50\right)$}   \\ \hline
  \multicolumn{1}{|c|}{\textbf{Circuit Depth$^\dagger$}}   & \multicolumn{1}{c|}{$397\ \left(\pm 76\right)$}   & \multicolumn{1}{c|}{$269\ \left(\pm
  50\right)$}   & \multicolumn{1}{c|}{$260\ \left(\pm 44\right)$}   \\ \hline
                                                         &                                                   &                                                   &                                                   \\ \hline
\multicolumn{1}{|c|}{\textbf{Parameter setting}}         & \multicolumn{3}{c|}{\begin{tabular}[c]{@{}c@{}}offline in simulation;\\ COBYQA optimizer~\cite{Ragonneau2023COBYQA}\end{tabular}}               \\ \hline
\multicolumn{1}{|c|}{\textbf{Shots per circuit}}         & \multicolumn{3}{c|}{$10,000$}                                                                                                                             \\ \hline
\multicolumn{1}{|c|}{\textbf{Bias parameter$^\ddagger$}} & \multicolumn{3}{c|}{\begin{tabular}[c]{@{}c@{}}$c=0.5$ for $i=0$;\\ $c\in\left\{0.1, 0.05\right\}$ for $i=1, 2, 3$;\\ $c\in\left\{0.05, 0.025\right\}$ for $i>3$\end{tabular}}                                                                                                                                  \\ \hline
\multicolumn{1}{|c|}{\textbf{Termination criteria}}      & \multicolumn{3}{c|}{\begin{tabular}[c]{@{}c@{}}Cost not improved \\ after 3 iterations\end{tabular}}                                                                                                                                  \\ \hline
\end{tabular}}
\caption{\label{tab:summary_implementation} Summary of implementation details of the presented experiments. Values formatted as $X\ \left(\pm \sigma\right)$ denote mean and $1$ standard deviation over all Hamiltonian instances/circuits.
\\
\footnotesize{$^\ast$percentage of the total Hamiltonian interactions}
\\
\footnotesize{$^\dagger$excluding virtual $RZ$ rotations}
\\
\footnotesize{$^\ddagger$index $i$ denotes ND-AWS iteration; two values of $c$ mean both were implemented and better solution was selected}
}
\end{table}

\begin{figure*}[t!]
\centering
\includegraphics[width=\textwidth]{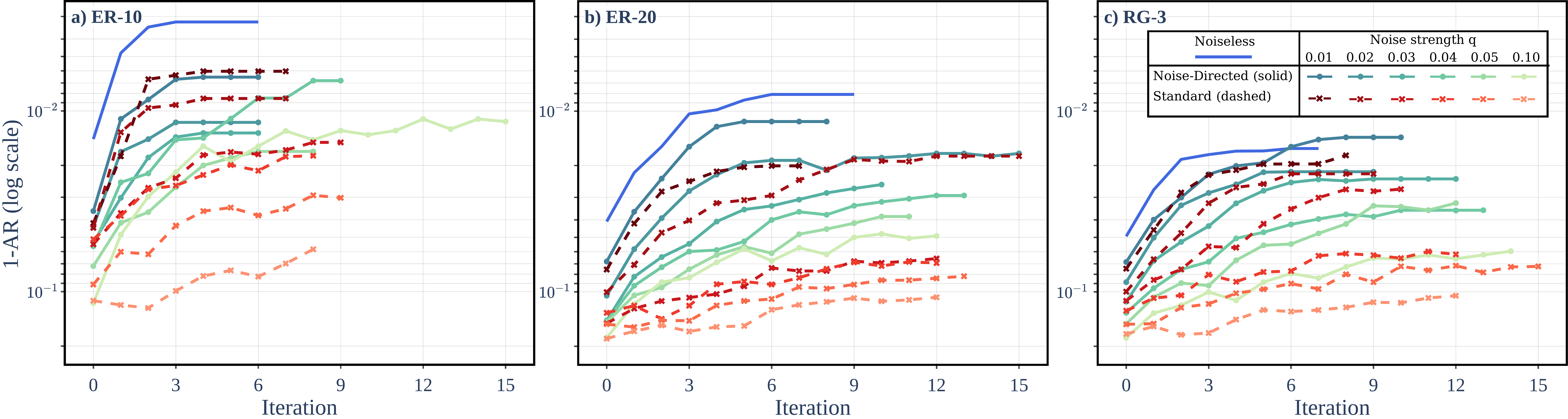}
\caption{\label{fig:sim:small_scale_performance}
Distance to the optimal solution, $1-AR$, of the best-found sample obtained in simulated runs of $\noq=20$-qubit random Erd\H{o}s-R\'enyi graphs with edge density $10\%$ (a) and $20\%$ (b), as well as 3-regular graphs (c).
The vertical axis is logarithmic and inverted, so that better solutions (lower $1-AR$) appear towards the top.
The blue solid line corresponds to a noiseless ND/Standard IWS implementation.
The solid lines with a greenish gradient correspond to the Noise-Directed WS ansatz with progressively stronger amplitude damping noise of strength $q$.
The dashed lines with a reddish gradient correspond to the Standard WS ansatz, without gauge transforms, as described in the text; the legend in (c), which applies to all panels, pairs the ND and Standard curves for each value of $q$.
Each data point is a mean over $10$ random Hamiltonian instances and $3$ independently-seeded solver runs per instance.
For clarity, we do not plot standard deviation, as there exists a huge spread over instances/runs, especially for the simulations with large noise strength, likely magnified due to small problem sizes and a small number of samples per run. 
The small $q$ lines are generally within run-to-run spread. 
} 
\end{figure*}

\subsection{Numerical proof of concept}\label{sec:numerics}\label{sec:numerics:small_scale_iws}

As a numerical proof of concept of the noise adaptivity of our approach, before turning to the experiments, we study $p=1$ ND-AWS with $H_\mathrm{PS}=H_C$ in small-scale $\noq=20$ qubit systems.
We optimize the $\beta$ and $\gamma$ angles with the expected-values simulator and perform sampling only from the optimized state.
Further supporting simulations -- approximate MPS simulations of $\noq=100$-qubit systems and a $500$-qubit study based on the efficient computation of expected values -- are presented in Appendix~\ref{app:more_numerics}.

We implement the full feedback loop for Hamiltonians $H_C$ with a connectivity built from random Erd\H{o}s-R\'enyi graphs with $10\%$ (ER-10) and $20\%$ (ER-20) edge probability; and 3-regular graphs (RG-3). 
We implement the protocol with $p=1$ WS-QAOA for $10$ random instances of each class, ER-10, ER-20, and RG-3, for $\noq=20$ qubits. 
Each instance is run $3$ times with different seeds, and below we report averages over both the instances and runs.
For such small problem sizes, sampling the ground state of the cost Hamiltonian with $p=1$ QAOA can happen since the entire search space only has $2^{20}\simeq 10^6$ states.
Therefore, at each iteration, we only draw $100$ samples from the optimized state.
The bias parameter starts at $c=0.5$ for iteration $0$, and is reduced to $c=0.1$ for iterations $1,2,3$, and to $c=0.05$ for further iterations.
The feedback loop terminates if $3$ consecutive iterations do not find a better solution, or as soon as the known optimal solution is sampled.

We implement both Standard and ND ansätze. 
In an ideal setting, both implementations give identical results, an equivalence broken by noise. 
We perform simulations with an uncorrelated, identical amplitude damping channel appended to each single-qubit gate in the circuit.
The Kraus operators of the single-qubit channel are
\begin{align}
    K_0 = \ketbra{0}{0}+\sqrt{1-q}\ \ketbra{1}{1}, && K_1 = \sqrt{q}\ \ketbra{0}{1} \ .
\end{align}
We perform simulations with varying amplitude damping strength $q \in \left\{0.01, 0.02, 0.03, 0.04, 0.05, 0.1\right\}$.

The results are presented in Fig.~\ref{fig:sim:small_scale_performance}.
Both Standard and ND ansätze are visibly affected by noise at the smallest strength of $q=0.01$, with RG-3 runs being the most robust. 
As noise strength grows, the AR decreases in all cases. 
Crucially, this effect is much stronger for the ansatz without the ND transformation.
In particular, for ER-10 and ER-20, the performance is similar for the ND ansatz with large noise of $q=0.1$ and the standard ansatz with much smaller noise of $q=0.03$.
The analogous crossover point is $q=0.04$ for RG-3. 
In general, the ND ansatz visibly tolerates higher amounts of noise in all cases.
For the largest noise strength $q=0.1$, we observe the ND algorithm to still achieve average ARs over $~\approx 0.98$ (ER-10) and $\approx 0.94$ (ER-20 and RG-3).

The quantitative differences between the ER Hamiltonian classes most likely come from the increased circuit depth required to implement ER-20 compared to ER-10. 
Interestingly, despite a smaller edge density, the RG-3 seems to perform the worst in both noisy and noiseless settings. 
We conclude that, on average, the ND ansatz highly outperforms its non-noise-aware counterpart in simulations when amplitude damping noise is present. 

\section{Experimental demonstration}\label{sec:exp}

\begin{figure}[t!]
\centering
\includegraphics[width=\columnwidth]{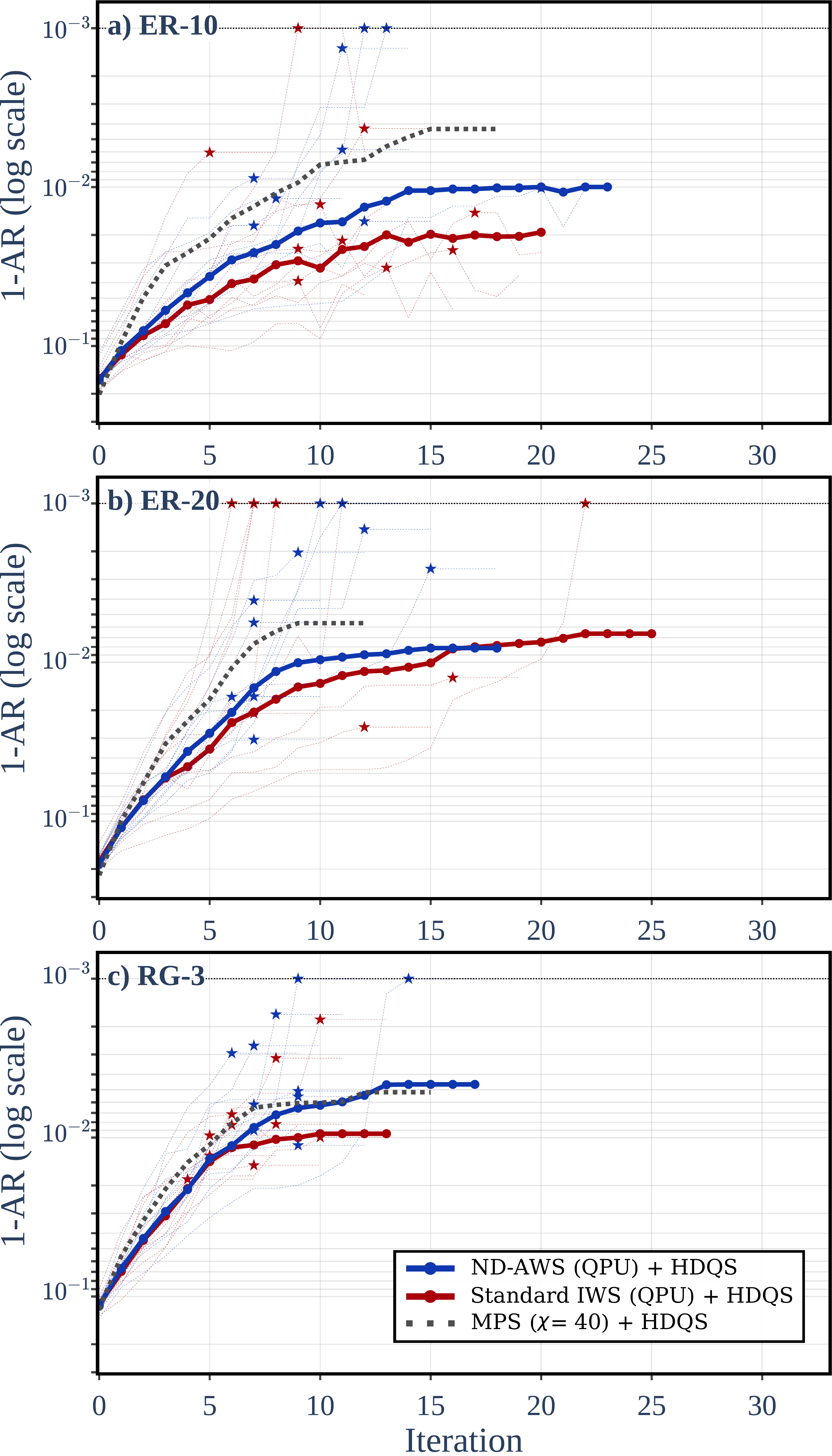}
\caption{\label{fig:exp:ndar_performance}
Distance to the optimal solution, $1-AR$ (logarithmic scale; inverted axis, as in Fig.~\ref{fig:sim:small_scale_performance}), of the best-found sample at each iteration of $p=1$ WS-QAOA implemented for $30$ Hamiltonians, ten for ER-10 (a), ten for ER-20 (b), and ten for RG-3 (c), with $\noq=100$ qubits on \texttt{ibm\_boston} with (ND-AWS, solid blue) and without (Standard, dashed red) the gauge transformations; the legend in (c) applies to all panels.
The results are augmented with $2$-local classical local search at each iteration (HDQS; see text description).
Optimal solutions ($AR=1$) are clipped to $10^{-3}$ (dotted horizontal line).
Each data point is the best of three independent runs.
Thick lines with circle markers correspond to the mean over all instances, while each thin dotted line is a different random Hamiltonian instance, with its best point indicated by a star symbol.
The grey dotted line is the mean over instances of an approximate MPS simulation of $p=1$ ND-AWS.}
\end{figure}

\subsection{Implementation details}

We implement ND-AWS for $10$ random instances of each class, ER-10, ER-20, and RG-3, on $100$ qubits.
Each instance is run $3$ times.
Below, we report the best out of the $3$ runs for each instance.

We implement $p=1$ ND-AWS with $25\%$ of the Hamiltonian interactions implemented in the phase separator for ER-10, $10\%$ for the denser ER-20; and $80\%$ for RG-3.
The interactions chosen for the PS operator are the relevant fraction of the \emph{largest-magnitude} interactions for the first $4$ iterations, and chosen at random afterwards.
At iteration $r=0$, we set $c=0.5$.
For the next iterations, we independently implement two values of the bias parameter and choose the better result as our candidate solution.
For iterations $r=1,2,3$, we use $c \in \left\{0.1, 0.05\right\}$, and for $r>3$, we use $c \in \left\{0.05, 0.025\right\}$ (recall Fig.~\ref{fig:illustration_IWS}).
This heuristic is motivated by numerical results presented in Appendix~\ref{app:numerics:large_scale_p1}.

The angles $\beta$ and $\gamma$ are optimized offline on classical hardware with an efficient $p=1$ simulation discussed in Appendix~\ref{app:numerics:large_scale_p1}. 
For each bias parameter $c$, we optimize the variational angles using the \texttt{COBYQA} optimizer~\cite{Ragonneau2023COBYQA} with basin hopping~\cite{wales1997basinhopping}, and at most $200$ cost function evaluations in total.
We then draw $s=10,000$ samples from the ansatz run on the QPU with the optimized angles.
This is followed by exploring the $2$-local Hamming-distance neighborhood of each sample and choosing the best result as our candidate solution at a given iteration.
We refer to this strategy as Hamming Distance Quadratic Search (HDQS) -- see Appendix~\ref{app:hdls_runs} for details and results without this classical post-processing.

\begin{table*}[t!]
\centering
\resizebox{\textwidth}{!}{%
\begin{tabular}{cclcccclcccclc}
\cline{1-2} \cline{4-7} \cline{9-12} \cline{14-14}
\multicolumn{1}{|l|}{\multirow{2}{*}{\textbf{Hamiltonian}}} & \multicolumn{1}{c|}{\multirow{2}{*}{\textbf{Ansatz}}} & \multicolumn{1}{l|}{} & \multicolumn{4}{c|}{\textbf{Iterations at convergence}}                                                                                                                         & \multicolumn{1}{l|}{} & \multicolumn{4}{c|}{\textbf{Best AR at convergence}}                                                                                                                               & \multicolumn{1}{l|}{} & \multicolumn{1}{c|}{\textbf{AR at iteration 10}} \\ \cline{4-7} \cline{9-12} \cline{14-14} 
\multicolumn{1}{|l|}{}                                      & \multicolumn{1}{c|}{}                                 & \multicolumn{1}{l|}{} & \multicolumn{1}{l|}{\textbf{min}} & \multicolumn{1}{l|}{\textbf{max}} & \multicolumn{1}{c|}{\textbf{median}} & \multicolumn{1}{c|}{\textbf{mean ($\pm$ SD)}}     & \multicolumn{1}{l|}{} & \multicolumn{1}{l|}{\textbf{min}} & \multicolumn{1}{l|}{\textbf{max}} & \multicolumn{1}{c|}{\textbf{median}} & \multicolumn{1}{c|}{\textbf{mean ($\pm$ SD)}}        & \multicolumn{1}{l|}{} & \multicolumn{1}{c|}{\textbf{mean ($\pm$ SD)}} \\ \cline{1-2} \cline{4-7} \cline{9-12} \cline{14-14} 
\multicolumn{1}{l}{}                                        & \multicolumn{1}{l}{}                                  &                       & \multicolumn{1}{l}{}              & \multicolumn{1}{l}{}              & \multicolumn{1}{l}{}                 & \multicolumn{1}{l}{}                              &                       & \multicolumn{1}{l}{}              & \multicolumn{1}{l}{}              & \multicolumn{1}{l}{}                 & \multicolumn{1}{l}{}                                 & & \\ \cline{1-2} \cline{4-7} \cline{9-12} \cline{14-14} 
\multicolumn{1}{|c|}{\multirow{3}{*}{ER-10}}                & \multicolumn{1}{c|}{Standard IWS}                         & \multicolumn{1}{l|}{} & \multicolumn{1}{c|}{$8$}           & \multicolumn{1}{c|}{$20$}           & \multicolumn{1}{c|}{$13.5$}            & \multicolumn{1}{c|}{$14.1\ \left(\pm 3.4\right)$} & \multicolumn{1}{c|}{} & \multicolumn{1}{c|}{$0.961$}         & \multicolumn{1}{c|}{$0.9998$}         & \multicolumn{1}{c|}{$0.977$}          & \multicolumn{1}{c|}{$0.981\ \left(\pm 0.012\right)$}   & \multicolumn{1}{l|}{} & \multicolumn{1}{c|}{$0.968\ \left(\pm 0.028\right)$} \\ \cline{2-2} \cline{4-7} \cline{9-12} \cline{14-14} 
\multicolumn{1}{|c|}{}                                      & \multicolumn{1}{c|}{\cellcolor{black!10}ND-AWS}                   & \multicolumn{1}{l|}{\cellcolor{black!10}} & \multicolumn{1}{c|}{\cellcolor{black!10}$10$}           & \multicolumn{1}{c|}{\cellcolor{black!10}$23$}           & \multicolumn{1}{c|}{\cellcolor{black!10}$14$}          & \multicolumn{1}{c|}{\cellcolor{black!10}$13.9\ \left(\pm 3.8\right)$} & \multicolumn{1}{c|}{\cellcolor{black!10}} & \multicolumn{1}{c|}{\cellcolor{black!10}$0.974$}         & \multicolumn{1}{c|}{\cellcolor{black!10}$1.0$}          & \multicolumn{1}{c|}{\cellcolor{black!10}$0.991$}          & \multicolumn{1}{c|}{\cellcolor{black!10}$0.990\ \left(\pm 0.008\right)$}   & \multicolumn{1}{l|}{\cellcolor{black!10}} & \multicolumn{1}{c|}{\cellcolor{black!10}$0.983\ \left(\pm 0.015\right)$} \\ \cline{2-2} \cline{4-7} \cline{9-12} \cline{14-14} 
\multicolumn{1}{|c|}{}                                      & \multicolumn{1}{c|}{MPS}                              & \multicolumn{1}{l|}{} & \multicolumn{1}{c|}{$6$}            & \multicolumn{1}{c|}{$18$}           & \multicolumn{1}{c|}{$9$}            & \multicolumn{1}{c|}{$10.6\ \left(\pm 4.0\right)$}   & \multicolumn{1}{l|}{} & \multicolumn{1}{c|}{$0.989$}         & \multicolumn{1}{c|}{$1.0$}          & \multicolumn{1}{c|}{$1.0$}          & \multicolumn{1}{c|}{$0.996\ \left(\pm 0.005\right)$}   & \multicolumn{1}{l|}{} & \multicolumn{1}{c|}{$0.993\ \left(\pm 0.010\right)$} \\ \cline{1-2} \cline{4-7} \cline{9-12} \cline{14-14} 
\multicolumn{1}{l}{}                                        & \multicolumn{1}{l}{}                                  &                       & \multicolumn{1}{l}{}              & \multicolumn{1}{l}{}              & \multicolumn{1}{l}{}                 & \multicolumn{1}{l}{}                              &                       & \multicolumn{1}{l}{}              & \multicolumn{1}{l}{}              & \multicolumn{1}{l}{}                 & \multicolumn{1}{l}{}                                 & & \\ \cline{1-2} \cline{4-7} \cline{9-12} \cline{14-14} 
\multicolumn{1}{|c|}{\multirow{3}{*}{ER-20}}                & \multicolumn{1}{c|}{Standard IWS}                         & \multicolumn{1}{l|}{} & \multicolumn{1}{c|}{$9$}           & \multicolumn{1}{c|}{$25$}           & \multicolumn{1}{c|}{$10.5$}            & \multicolumn{1}{c|}{$13.3\ \left(\pm 4.9\right)$} & \multicolumn{1}{c|}{} & \multicolumn{1}{c|}{$0.974$}         & \multicolumn{1}{c|}{$1.0$}        & \multicolumn{1}{c|}{$1.0$}          & \multicolumn{1}{c|}{$0.994\ \left(\pm 0.009\right)$}   & \multicolumn{1}{l|}{} & \multicolumn{1}{c|}{$0.987\ \left(\pm 0.016\right)$} \\ \cline{2-2} \cline{4-7} \cline{9-12} \cline{14-14} 
\multicolumn{1}{|c|}{}                                      & \multicolumn{1}{c|}{\cellcolor{black!10}ND-AWS}                   & \multicolumn{1}{l|}{\cellcolor{black!10}} & \multicolumn{1}{c|}{\cellcolor{black!10}$9$}            & \multicolumn{1}{c|}{\cellcolor{black!10}$18$}           & \multicolumn{1}{c|}{\cellcolor{black!10}$11$}            & \multicolumn{1}{c|}{\cellcolor{black!10}$12.1\ \left(\pm 2.7\right)$} & \multicolumn{1}{c|}{\cellcolor{black!10}} & \multicolumn{1}{c|}{\cellcolor{black!10}$0.969$}         & \multicolumn{1}{c|}{\cellcolor{black!10}$1.0$}        & \multicolumn{1}{c|}{\cellcolor{black!10}$0.997$}          & \multicolumn{1}{c|}{\cellcolor{black!10}$0.992\ \left(\pm 0.010\right)$}   & \multicolumn{1}{l|}{\cellcolor{black!10}} & \multicolumn{1}{c|}{\cellcolor{black!10}$0.990\ \left(\pm 0.009\right)$} \\ \cline{2-2} \cline{4-7} \cline{9-12} \cline{14-14} 
\multicolumn{1}{|c|}{}                                      & \multicolumn{1}{c|}{MPS}                              & \multicolumn{1}{l|}{} & \multicolumn{1}{c|}{$6$}            & \multicolumn{1}{c|}{$12$}           & \multicolumn{1}{c|}{$10$}          & \multicolumn{1}{c|}{$9.7\ \left(\pm 1.7\right)$} & \multicolumn{1}{l|}{} & \multicolumn{1}{c|}{$0.976$}         & \multicolumn{1}{c|}{$1.0$}          & \multicolumn{1}{c|}{$0.998$}          & \multicolumn{1}{c|}{$0.995\ \left(\pm 0.007\right)$}   & \multicolumn{1}{l|}{} & \multicolumn{1}{c|}{$0.995\ \left(\pm 0.007\right)$} \\ \cline{1-2} \cline{4-7} \cline{9-12} \cline{14-14} 
\multicolumn{1}{l}{}                                        & \multicolumn{1}{l}{}                                  &                       & \multicolumn{1}{l}{}              & \multicolumn{1}{l}{}              & \multicolumn{1}{l}{}                 & \multicolumn{1}{l}{}                              &                       & \multicolumn{1}{l}{}              & \multicolumn{1}{l}{}              & \multicolumn{1}{l}{}                 & \multicolumn{1}{l}{}                                 & & \\ \cline{1-2} \cline{4-7} \cline{9-12} \cline{14-14} 
\multicolumn{1}{|c|}{\multirow{3}{*}{RG-3}}                 & \multicolumn{1}{c|}{Standard IWS}                         & \multicolumn{1}{l|}{} & \multicolumn{1}{c|}{$7$}           & \multicolumn{1}{c|}{$13$}           & \multicolumn{1}{c|}{$9.5$}            & \multicolumn{1}{c|}{$9.9\ \left(\pm 2.0\right)$} & \multicolumn{1}{c|}{} & \multicolumn{1}{c|}{$0.982$}         & \multicolumn{1}{c|}{$0.998$}        & \multicolumn{1}{c|}{$0.991$}          & \multicolumn{1}{c|}{$0.991\ \left(\pm 0.005\right)$}   & \multicolumn{1}{l|}{} & \multicolumn{1}{c|}{$0.991\ \left(\pm 0.005\right)$} \\ \cline{2-2} \cline{4-7} \cline{9-12} \cline{14-14} 
\multicolumn{1}{|c|}{}                                      & \multicolumn{1}{c|}{\cellcolor{black!10}ND-AWS}                   & \multicolumn{1}{l|}{\cellcolor{black!10}} & \multicolumn{1}{c|}{\cellcolor{black!10}$9$}            & \multicolumn{1}{c|}{\cellcolor{black!10}$17$}           & \multicolumn{1}{c|}{\cellcolor{black!10}$11.5$}            & \multicolumn{1}{c|}{\cellcolor{black!10}$11.5\ \left(\pm 2.1\right)$} & \multicolumn{1}{c|}{\cellcolor{black!10}} & \multicolumn{1}{c|}{\cellcolor{black!10}$0.989$}         & \multicolumn{1}{c|}{\cellcolor{black!10}$1.0$}        & \multicolumn{1}{c|}{\cellcolor{black!10}$0.996$}          & \multicolumn{1}{c|}{\cellcolor{black!10}$0.996\ \left(\pm 0.004\right)$}   & \multicolumn{1}{l|}{\cellcolor{black!10}} & \multicolumn{1}{c|}{\cellcolor{black!10}$0.994\ \left(\pm 0.005\right)$} \\ \cline{2-2} \cline{4-7} \cline{9-12} \cline{14-14} 
\multicolumn{1}{|c|}{}                                      & \multicolumn{1}{c|}{MPS}                              & \multicolumn{1}{l|}{} & \multicolumn{1}{c|}{$7$}           & \multicolumn{1}{c|}{$15$}           & \multicolumn{1}{c|}{$9.5$}          & \multicolumn{1}{c|}{$9.8\ \left(\pm 2.0\right)$} & \multicolumn{1}{l|}{} & \multicolumn{1}{c|}{$0.987$}        & \multicolumn{1}{c|}{$0.9998$}        & \multicolumn{1}{c|}{$0.996$}          & \multicolumn{1}{c|}{$0.995\ \left(\pm 0.004\right)$} & \multicolumn{1}{l|}{} & \multicolumn{1}{c|}{$0.994\ \left(\pm 0.004\right)$} \\ \cline{1-2} \cline{4-7} \cline{9-12} \cline{14-14} 
\end{tabular}}
\caption{\label{tab:exp:ndar_performance}
Table summarizing the data from Fig.~\ref{fig:exp:ndar_performance} for the iterative implementation of Standard and Noise-Directed ansätze (QPU), as well as MPS simulations, augmented with Hamming Distance Quadratic Search. 
For a fixed Hamiltonian class and ansatz, the indicated functions are applied over random Hamiltonian instances.
The last column reports the mean ($\pm$ SD over instances) of the best-found AR at iteration $10$; for runs that converged before iteration $10$, the final value is carried over.}
\end{table*}

Between ND-AWS iterations, we use a greedy strategy that always selects the solution if it is better than previously found ones. 
The execution terminates if three consecutive iterations do not improve the best found solution.

We implement all of our experiments on $100$ qubits of the \texttt{ibm\_boston} QPU.
The Time-Block phase separator Hamiltonian is routed to the device's topology using the \texttt{SABRE} algorithm~\cite{li2019sabre,zou2024lightsabre}, available in the \texttt{Qiskit} library~\cite{qiskit2024}. 
For each run, we reject some of the noisier qubits when defining the allowed gate set for the transpiler. Table~\ref{tab:summary_implementation} summarizes the implementation parameters while Appendix~\ref{app:hardware_implementation} presents more details.

\subsection{Results}

To study the noise-adaptivity of the noise-directed ansatz, we implement each experiment with (ND-AWS) and without (Standard IWS) the gauge transformation that aligns the best-found solution with $\ket{0\dots 0}$.
Without the gauge transformation, the procedure outlined in Section~\ref{sec:ansatz} most closely resembles the approach of Ref.~\cite{lopez2025non}; and the $c$ parameter is defined for each qubit $i$ as $c_i \rightarrow c$ if $y_i=0$ and $c_i \rightarrow 1-c$ if $y_i=1$.

With the gauge transformations, we find solutions with AR between $0.974$--$1.0$, $0.969$--$1.0$ and $0.989$--$1.0$, for the ER-10, ER-20, and RG-3 Hamiltonians, respectively, see Table~\ref{tab:exp:ndar_performance} and Fig.~\ref{fig:exp:ndar_performance}.
The convergence is achieved in $10$--$23$, $9$--$18$, and $9$--$17$ iterations, meaning that $2.1\cdot10^5$--$4.7\cdot10^5$, $1.9\cdot10^5$--$3.7\cdot10^5$, and $1.9\cdot10^5$--$3.5\cdot10^5$ samples were generated on a QPU to find the reported solutions~\footnote{Please note that we label iterations starting from $0$. As explained in the text, $0$th iteration corresponds to standard (Time-Block) QAOA with $c=0.5$. Each consecutive iteration is implemented with $2$ distinct values of $c$, hence the total number of samples generated until iteration $r$ is $10^4\left(2r+1\right)$.}, for the ER-10, ER-20, and RG-3 Hamiltonians, respectively.
For ER-10, without the gauge transformation, we converge in slightly fewer iterations, i.e., $8$--$20$, but reach lower ARs, i.e., $0.961$--$0.9998$, than with the gauge transformations.
For ER-20, the ansatz without the gauge transformations gives solutions of similar quality (AR $0.974$--$1.0$) to the ansatz with the gauge transformations, but at the cost of more iterations ($9$--$25$).
We note that in this case, Standard WS anastz performed particularly well, finding an optimal solution for more than half of the instances (median $1.0$ in Table~\ref{tab:exp:ndar_performance}). 
For RG-3, similarly to ER-10, the ansatz without gauge transformations is faster to converge ($7$--$13$) to solutions of lower quality (AR $0.982$--$0.998$).

For ER-10 and RG-3, the noise-directed ansatz results in smoother convergence profiles, with the AR being less rugged and less spread out (in the lower-performance direction) than without the gauge transformation; see thin dotted lines in Fig.~\ref{fig:exp:ndar_performance}.
This is particularly visible for the ER-10 Hamiltonians. 
Here, the non-noise-aware ansatz has instances with large AR drops between iterations, i.e., the ansatz is prone to diverging into higher-energy (worse-quality) regions. 
We conclude that the ND ansatz generally either improves (ER-10 and RG-3) or gives similar (ER-20) performance relative to the non-ND ansatz at no additional circuit complexity.
Small-scale simulations with a controlled amplitude-damping noise strength corroborate this noise-adaptivity mechanism; see Section~\ref{sec:numerics:small_scale_iws}.

We also perform approximate Matrix Product State (MPS) simulations at a bond dimension of $\chi=40$ of $p=1$ ND-AWS with the full phase separator Hamiltonian, i.e., no Time-Block ansatz (see Appendix~\ref{sec:numerics:mps} for details of these simulations).
The MPS results outperform the QPU ones---offering a higher/similar quality with a faster convergence, see the thick grey dotted lines in Fig.~\ref{fig:exp:ndar_performance}; indicating future hardware improvements will lead to increased performance in quantum optimization.

\section{Discussion} \label{sec:discussion}

We introduce Noise-Directed Adaptive Warm-Starting (ND-AWS), a new approach to quantum optimization that combines ideas from Warm-Starting~\cite{egger2021warm}, Noise-Directed Adaptive Remapping~\cite{maciejewski2024ndar}, and classical search algorithms.
To make the QPU implementation feasible, we employ the Time-Block ansatz~\cite{maciejewski2023design} to implement a phase separator with only part of the Hamiltonian interactions.
To counteract energy relaxation, we bias the optimization towards the physical $\ket{0\dots 0}$ state, and employ bitflip gauge transformations to make $\ket{0\dots 0}$ logically equivalent to the best candidate solution at each iteration.

We test ND-AWS experimentally on $100$ qubits of \texttt{ibm\_boston} on a total of $30$ random Hamiltonian instances, and find high performance using QPU-obtained samples augmented with greedy, local classical post-processing.
Crucially, the gauge transformations generally increase performance in practice when compared against a non-gauge-transformed variant of Iterative Warm-Starting.
To the best of our knowledge, considering the scale and complexity of the implemented Hamiltonians, see Table~\ref{tab:summary_implementation}, the presented results are among the demonstrations of quantum applied optimization with the highest quality at scale to date~\cite{abbas2023quantum,Pelofske2024ScalingQAOA,wang2025linearchainQAOA, mohseni2026copulaQAOA, montanezbarrera2025evaluatingperformancequantumprocessing, montanezbarrera2025LinearRampQAOA, rava2025benchmarkingneutralatombasedquantum}.
Noiseless tensor network simulations (Appendix~\ref{sec:numerics:mps}) suggest that improving the hardware might further enhance performance.

Our work opens multiple possible research directions.
First, ND-AWS is intentionally simple so that it may serve as a conceptual framework, rather than a fully optimized solver.
In addition to improving the circuit ansatz structure and depth, improved results could be obtained by modifying some of the ND-AWS subroutines, such as including (i) a tree-like search over multiple best solutions as opposed to just a single one, similar to NDAR~\cite{maciejewski2024multilevel}, (ii) adaptive annealing schedules for bias parameters, (iii) optimization of angles directly on hardware instead of transferring angles from numerical simulations~\cite{maciejewski2023design, maciejewski2024ndar, Cheng2024DARBO}, and adding gate reordering as a new categorical parameter~\cite{maciejewski2023design,maciejewski2024multilevel}, (iv) improved Hamiltonian-to-device mapping beyond the \texttt{SABRE} algorithm~\cite{matsuo2023sat,kotil2023Routing,vcepaite2025quantum}, (v) sampling-compatible error mitigation and suppression methods~\cite{Quek2024exponentially,viola1998dynamical,Ezzell2023DDReview,Wallman2016RCoriginal,Hashim2021RCscalable,Nation2021M3REM}, and (vi) combining the QPU solver with a classical solver in the feedback loop.

Our algorithm builds off a
wealth of classical
approaches that
incorporate repeated refinements of solutions, often through iterated local searches~\cite{FeoResende1995, AartsLenstra2003, hoos2015stochastic, AartsLenstra2003, LourencoMartinStutzle2003}.
Many of these approaches may be directly leveraged to derive more sophisticated versions of our algorithm.
For example, to improve experiments presented in Fig.~\ref{fig:exp:ndar_performance}, we augmented each iteration with classical local search -- in Appendix~\ref{app:hdls_runs}, we demonstrate that experiments without this strategy were generally of lower quality and slower convergence.

Furthermore, ND-AWS is compatible with \emph{quantum-enhanced} optimization, where quantum samples help improve classical solvers -- a dual setting to the warm-start considered here.
Examples include Quantum Relax and Round~\cite{Dupont2024QRR,maciejewski2024multilevel}, its generalized version called Quantum Preconditioning~\cite{dupont2025optimization}, and the warm-start approach to improve a classical TABU search~\cite{vcepaite2025quantum, Palubeckis2004TABU}.
It would be interesting to study how ND-AWS may help in such settings.

We demonstrate our algorithm on a superconducting transmon device.
Here, amplitude damping is often present as seen, e.g., in previous work on NDAR~\cite{maciejewski2024ndar,maciejewski2024multilevel,tam2025enhancingNDAR}. 
One could further test ND-AWS on different quantum computing platforms to identify which technology benefits the most from it. 
Indeed, damping and relaxation mechanisms are common across physical architectures, from spontaneous emission and decay in trapped ions and neutral atom systems~\cite{ringbauer2025verifiable, wu2022erasure} to relaxation of spin qubits~\cite{lawrie2020spin}, among others.
For example, ionic or atomic platforms are typically orders of magnitude slower than superconducting ones, while exhibiting lower levels of noise and better connectivity. 
This means that different Warm-Start ansatz using different ND-AWS hyperparameters may be suitable for such devices. 
Similarly, incorporating parameter concentration/parameter transfer strategies~\cite{galda2021transferabilityoptimalqaoaparameters,galda2023similaritybasedparametertransferabilityquantum,Shaydulin2023ParameterTransfer,braydwood2026qaoa, guo2026setting} for WS ansätze could help alleviate speed limitations generally. 
Beyond qubits, it would be interesting to generalize the protocol to bosonic and qudit architectures~\cite{wang2020qudits,su2021construction,Blok2021QutritScrambling,chi2022programmable,Ringbauer2022UniversalQudit,roy2023two,denys2023,nguyen2024empowering,bornman2025benchmarking,kim2025ultracoherent,venturelli2025near} and alternative technologies like p-dit Ising machines~\cite{duffee2025p}. 

In summary, recent hardware progress allows researchers to test new quantum heuristics for combinatorial optimization at scale. 
At the same time, system noise remains a serious obstacle, especially for the foreseeable future. Our approach seeks to leverage the effects of noise where possible, rather than expend resources fighting it. To this end,
we introduce ND-AWS, combining ideas from warm-starting, NDAR, and classical algorithms, toward obtaining effective quantum heuristics today. 
As quantum computing technologies continue to progress and improved quantum circuits can be reliably deployed, we anticipate further performant heuristics to be developed, tested, 
refined, and 
benchmarked~\cite{abbas2023quantum,koch2025decathlon}. 

\section*{Data and Code availability}
The Python code implementing ND-AWS will be made publicly available in the GitHub repository \href{https://github.com/usra-riacs/quantum-approximate-optimization}{github.com/usra-riacs/quantum-approximate-optimization}~\cite{quapopt_repo}. 
The raw experimental data (measurement bitstrings from all QPU runs) cannot be made publicly available because of their size (dozens of gigabytes); they are available from the corresponding author upon reasonable request.

\section*{Acknowledgments}
This work was supported as a part of NCCR SPIN, a National Centre of Competence in Research, funded by the Swiss National Science Foundation (Grant No. 225153).
This work was supported by the Hartree National Centre for Digital Innovation, a collaboration between the Science and Technology Facilities Council and IBM.
DV, FBM and SH acknowledge AFRL Contract No. FA8750-25-C-B0040 and useful technical exchanges with Daniel Koch.
FBM is grateful to Bao Gia Bach and Ilya Safro for useful discussions on various stages of this project.
GP and OW would like to thank Stefano Mensa for his advice and technical discussions.

\bibliographystyle{apsrev4-2}
\bibliography{bibliography}

\clearpage

\onecolumngrid
\appendix

\begin{center}
{\large Supplementary Materials for\\ Quantum Approximate Optimization\\ via Noise-Directed Adaptive Warm-Starting}
\end{center}

\section{Warm-Starting Quantum Approximate Optimization}\label{app:warm_starting_qaoa}

In the main text, we briefly described our ansatz as a combination of Time-Block~\cite{maciejewski2023design} and Warm-Start~\cite{egger2021warm} ansätze with gauge-transformation ideas from Noise-Directed Adaptive Remapping~\cite{maciejewski2024ndar}.
For completeness, here we provide a more detailed description of these three elements. 

\subsection{QAOA}

Consider a diagonal cost Hamiltonian 
\begin{align}\label{eq:cost_hamiltonian}
    H_C = \sum_{i} h_{i} Z_i + \sum_{i<j} J_{ij} Z_iZ_j +\dots
\end{align}
where $Z_i$ is the Pauli $Z$ operator acting on qubit $i$, and $J_{ij} \in \mathbb{R}$ is the interaction strength between qubits $i,j$, while $h_i$ is a local field magnitude.
A wide variety of challenging optimization problems map to Hamiltonians of this form~\cite{lucas2014ising,hadfield2021representation}. 
We focus on two-local problems. 
Nevertheless, most of our observations (including iterative warm-starting) apply to higher-order Hamiltonians.
An $n$-qubit QAOA circuit~\cite{farhi2014quantum,hadfield2019quantum} of depth $p$ and parameters $\vec{\gamma},\vec{\beta}\in\mathbb{R}^p$ is given by
\begin{align}\label{eq:generic_qaoa_circuit}
U\left(H_C, H_M;\vec{\gamma},\vec{\beta}\right) := \prod_{l=p}^{1} U_{M}\left(H_M; \beta_l\right)U_{P}\left(H_{P}; \gamma_l\right), 
\end{align}
with a Phase Separator (PS)
\begin{align}
   U_{P}\left(H_{P};\gamma_l\right) = \exp\left(-i\gamma_l H_{P}\right) 
\end{align}
and Mixer operator
\begin{align}
    U_{M} \left(H_M; \beta_k\right) := \exp\left(-i\beta_k H_M\right)\ .
\end{align}
In the original variant~\cite{farhi2014quantum},
    $H_M = \sum_{j=0}^{n-1} X_j$,
$H_{PS} = H_{C}$, and the
operator $U\left(H_C, H_M;\vec{\gamma},\vec{\beta}\right) $ is applied to the standard initial state 
$\ket{s}=\ket{+}^{\otimes n}$, where $\ket{+} = \frac{1}{\sqrt{2}}\left(\ket{0}+\ket{1}\right)$.

\subsection{Warm-Start QAOA}

In Ref.~\cite{egger2021warm}, the authors introduced a modification of the original QAOA ansatz called Warm-Start QAOA (WS-QAOA).
WS-QAOA changes both the initial state and the mixer operator to bias the optimization towards a solution obtained via solvers that perform continuous-valued relaxation of the original binary-valued problem.

Consider an output of such a relaxation-based algorithm given as a vector $\mathbf{c} = \left(c_0, c_1, \dots, c_{n-1}\right)$, where $c_{i}\in \left[0,1\right]$.
In standard QAOA, the initial state for each qubit is equal superposition of $\ket{0}$ and $\ket{1}$ states.
In WS-QAOA, the initial state for the optimization is instead
\begin{align}\label{eq:app:ws_ininital_state}
   \ket{\mathbf{c}} =\bigotimes_{i=0}^{n-1} \ket{c_i} \ ,
\end{align}
where, for qubit $i$ we have
\begin{align}\label{eq:app:ws_ininital_state_one_qubit}
\ket{c_i} \coloneqq R_{Y}\left(2\arcsin(\sqrt{c_i})\right)\ket{0} = \sqrt{1-c_i}\ket{0} + \sqrt{c_i} \ket{1} \ .
\end{align}
To ensure that $\ket{\mathbf{c}}$ is the ground state of the negated mixer Hamiltonian $-H_{M}$, we now modify it to be
\begin{align}\label{eq:app:ws_mixer}
    H_{M} = \sum_{i} H_i^{M}\ ,
\end{align}
with
\begin{align}\label{eq:app:ws_mixer_one_qubit}
    H_{i}^{M} = \left(2\sqrt{c_i\left(1-c_i\right)}\right)X_i+\left(1-2c_i\right)Z_i \ .
\end{align}
Similarly to standard QAOA, the initial state is chosen as the ground state of the mixer, to preserve asymptotic depth performance guarantees related to the adiabatic theorem.

\subsection{Biasing WS-QAOA towards a single solution}
The initial WS ansatz biases the QAOA optimization towards the output $\mathbf{c}$ of a continuously-relaxed classical solver. 
By contrast, here we bias the optimization towards a binary solution (bitstring) $\mathbf{x}$ that can be an output of an arbitrary combinatorial optimization solver (including quantum optimizers).
To this aim, we generalize the regularization method from Ref.~\cite{egger2021warm}, by constructing a vector $\mathbf{c}$ defined via a set of functions $\left\{f_i|f_i:\left\{0,1\right\}\rightarrow\left[0,1\right]\right\}_{i=0}^{n-1}$ as
\begin{align}\label{eq:bias_vector_definition}
    \mathbf{c} =     \left(f_0\left(x_0\right),f_1\left(x_1\right), \dots, f_{n-1}\left(x_{n-1}\right)\right) \ .
\end{align}
To meaningfully introduce a bias, we require
\begin{align}\label{eq:bias_function_definition}
    f_i\left(x_i\right) =  \begin{cases}
    \leq \frac{1}{2}-\varepsilon & \text{if } x_i = 0 \\
    > \frac{1}{2}+\varepsilon & \text{if } x_i = 1 
\end{cases} \ .
\end{align}
A stronger bias is achieved for larger values of $\varepsilon$.
A simple example of $\left\{f_i\right\}$ that fulfills the above conditions is
\begin{align}\label{eq:bias_function_example}
    f_i\left(x_i\right) =  \begin{cases}
    t_i & \text{if } x_i = 0 \\
    1-t_i & \text{if } x_i = 1 
\end{cases} \ ,
\end{align}
where $t_i \in \left[0,\frac{1}{2}\right]$.
For $t_i=\frac{1}{2}$, we recover the original QAOA setting with no bias and equal superposition over both solutions (computational-basis states). 
Decreasing $t_i$ corresponds to biasing the optimization towards $x_i$ on qubit $i$, with the extreme case of $t_i=0$ corresponding to fixing the state of qubit $i$ to $x_i$ (in practice, one could simply construct a reduced $\left(n-1\right)$-qubit Hamiltonian instead). 
Note that setting all $t_i$ to $0$ corresponds to preparing exactly the solution $\mathbf{x}$ as the input state, and the WS ansatz becomes logically trivial (this is because the input state becomes $\ket{\mathbf{x}}$ and the mixer Hamiltonian becomes diagonal in the computational basis, so the whole circuit just adds a global phase to the input state).
Increasing $t_i$ away from $0$ can be conceptually viewed as increasing the support of the input state over computational basis states further away (in Hamming distance) from the input bitstring $\mathbf{x}$, allowing for larger-radius search over problem solutions.

\subsection{Gauge-transformed Noise-Directed WS-QAOA}\label{app:theory:gauges}

\subsubsection{Bitflip transforms} 
Consider the unitary ``bitflip" operator $P_{\y} = \bigotimes_{i=0}^{\noq-1}X_{i}^{y_i}$ that acts to flip the $|0\rangle$, $|1\rangle$ basis states as specified by the bitstring $\y\in\{0,1\}^\noq$.
This change-of-basis can be applied to $H$ 
by incorporating the change of signs to the weights of $h_i$ and $J_{i,j}$ as 
$H \rightarrow H^{\y}$ with 
\begin{align}\label{eq:app:gauge_transformation_local}
    H^{\y} 
    &= P_{\y}HP_{\y}\\
    &= \sum_{i} \left(-1\right)^{y_i}h_{i} Z_i 
    + \sum_{i<j} \left(-1\right)^{y_i+y_j}J_{i,j} Z_iZ_j +\dots. \nonumber
\end{align}
This transformation preserves cost Hamiltonian eigenvalues, with eigenvectors (candidate problem solutions) permuted~under $P_{\y}$. 
In particular, under this transformation the $|0\dots0\rangle$ state is mapped to~$|y_0\dots y_{n-1}\rangle$.
Moreover, for QAOA circuits, it is very simple to implement. 
As seen from Eq.~\eqref{eq:app:gauge_transformation_local}, it requires only changing signs in the respective gates. 
The measurement basis change is done in post-processing.

\subsubsection{Gauge transformations of WS-QAOA circuits}\label{app:qaoa_symmetries}

Now we will show that the WS-QAOA ansatz with bias towards $\ket{0 \dots 0}$ and gauge-transformed phase separator $H_{C}^{\mathbf{x}}$ is, in a noiseless setting, equivalent to implementing WS-QAOA biased towards $\ket{\mathbf{x}}$ with standard phase separator $H_{C}$.
Here we prove a more general statement that the WS-QAOA ansatz biased towards $\mathbf{x}$ with a gauge-transformed phase separator $H_{C}^{\mathbf{y}}$ is equivalent to implementing the WS-QAOA ansatz biased towards a flipped bitstring $\ket{\mathbf{y}\oplus\mathbf{x}}$ with the original phase separator $H_{C}$.
This general case recovers biasing towards $\ket{0 \dots 0}$ if we set $\ket{\mathbf{y}}=\ket{\mathbf{x}}$.
In what follows, we assume that the initial WS-QAOA ansatz is biased towards some state $\mathbf{x}$.
Moreover, we assume that the cost Hamiltonian and Phase Separator Hamiltonian are the same -- the proof trivially generalizes to the alternative case (including Time-Block ansatz).
We will use the following properties of Pauli matrices: $P_{\mathbf{y}}^2=\mathbb{I}$, $P_{\mathbf{y}}=P_{\mathbf{y}}^{\dagger}$, $P_{\mathbf{y}}XP_{\mathbf{y}}=X$, and $P_{\mathbf{y}}Z_jP_{\mathbf{y}} = \left(-1\right)^{y_j}Z_j$.

First, recall that a single term in the WS-QAOA mixer Hamiltonian biased towards $\ket{\mathbf{x}}$ is given by
\begin{align}\label{eq:app:ws_mixer_one_qubit_general}
    H_{i}^{M} = a_i X_i+b_iZ_i \ ,
\end{align}
with 
\begin{align}
  a_i =   2\sqrt{c_i\left(1-c_i\right)}, && b_i = 1-2c_i
\end{align}
and the corresponding input state on qubit $i$ by
\begin{align}\label{eq:app:ws_ininital_state_one_qubit2}
\ket{c_i} \coloneqq R_{Y}\left(2\arcsin(\sqrt{c_i})\right)\ket{0} = \sqrt{1-c_i}\ket{0} + \sqrt{c_i} \ket{1} \ .
\end{align}
In the above, $c_i=f_i\left(x_i\right)$ implicitly depends on $\ket{\mathbf{x}}$, and to have bias we require that $c_i<\frac{1}{2}$ if $x_i=0$ and $c_i>\frac{1}{2}$ if $x_i=1$, recall Eq.~\eqref{eq:bias_function_definition}.
To implement a WS-QAOA biased towards a flipped state $P_{\mathbf{y}}\ket{\mathbf{x}} = \ket{\y\oplus\mathbf{x}}$, we notice that, according to Eq.~\eqref{eq:bias_function_definition}, flipping bias of qubit $i$ from $x_i$ to $1-x_i$ corresponds to transformation $c_i \rightarrow 1-c_i$, which changes the input state to 
\begin{align}
  \ket{c_i} \rightarrow \sqrt{c_i} \ket{0} + \sqrt{1-c_i} \ket{1} \  = X \ket{c_i},
\end{align}
and the mixer to
\begin{align}\label{eq:app:mixer_flip}
   H_i^M \rightarrow a_i X_i - b_iZ_i  = X_i H_i^M X_i \ .
\end{align}

Recall notation 
\begin{equation}
    H^{\mathbf{y}}_C = P_{\mathbf{y}} H_C P_{\mathbf{y}} \ .
\end{equation}
Now consider the gauge-transformed WS-QAOA unitary
\begin{equation}
\begin{split}
 U\left(H_{C}^{\mathbf{y}}, H_{M};\vec{\gamma},\vec{\beta}\right) = \prod_{l=p}^{1} U_{M}\left(H_M; \beta_l\right)U_{PS}\left(H_C^{\mathbf{y}}; \gamma_l\right) =  
  \prod_{l=p}^{1} U_{M}\left(H_M; \beta_l\right)P_{\mathbf{y}}U_{PS}\left(H_C; \gamma_l\right)P_{\mathbf{y}} = \\
    \prod_{l=p}^{1} P_{\mathbf{y}} P_{\mathbf{y}} U_{M}\left(H_M; \beta_l\right)P_{\mathbf{y}}U_{PS}\left(H_C; \gamma_l\right)P_{\mathbf{y}} =     \prod_{l=p}^{1} P_{\mathbf{y}}  U_{M}\left(H_M^{\mathbf{y}}; \beta_l\right)U_{PS}\left(H_C; \gamma_l\right)P_{\mathbf{y}}  \ ,
\end{split}
\end{equation}
where we denoted $H_M^{\mathbf{y}}$ as the WS-QAOA Mixer Hamiltonian biased towards $\ket{\mathbf{x}\oplus \mathbf{y}}$ state, in the sense described previously (dependence on $\x$ is implicit for ease of notation).

Now consider computational-basis state $\ket{\mathbf{b}}$ and corresponding amplitude of the gauge-transformed WS-QAOA state vector:
\begin{equation}
\begin{split}
      \bra{\mathbf{b}\oplus \mathbf{y}} U\left(H_{C}^{\mathbf{y}}, H_{M};\vec{\gamma},\vec{\beta}\right)\ket{\mathbf{c}} =  \bra{\mathbf{b}\oplus \mathbf{y}}  \left(\prod_{l=p}^{1} P_{\mathbf{y}}  U_{M}\left(H_M^{\mathbf{y}}; \beta_l\right)U_{PS}\left(H_C; \gamma_l\right)P_{\mathbf{y}} \right) \ket{\mathbf{c}} = \\ 
      \bra{\mathbf{b}\oplus \mathbf{y}} P_{\mathbf{y}}  \left(\prod_{l=p}^{1}  U_{M}\left(H_M^{\mathbf{y}}; \beta_l\right)U_{PS}\left(H_C; \gamma_l\right)\right) P_{\mathbf{y}} \ket{\mathbf{c}} = 
        \bra{\mathbf{b}} U\left(H_{C}, H_{M}^{\mathbf{y}};\vec{\gamma},\vec{\beta}\right)\ket{\mathbf{c}^{\mathbf{y}}} \ ,
\end{split}
\end{equation}
where we denoted $\ket{c^{\mathbf{y}}}$ as the WS-QAOA input state biased towards $\ket{\mathbf{x}\oplus \mathbf{y}}$ state.
In the above, $\bra{\mathbf{b}\oplus \y}$ comes from a change-of-basis operation necessary to measure $H^{\y}_{C}$ instead of $H_C$.
Thus, we have shown that implementing WS-QAOA biased towards $\mathbf{x}$ with gauge-transformed cost $H_{C}^{\mathbf{y}}$ (LHS above; note that changing the cost Hamiltonian in quantum optimization involves both changing the phase separator and the measurement basis) is equivalent to implementing WS-QAOA biased towards $\ket{\mathbf{x}\oplus \mathbf{y}}$ with the original cost $H_{C}$ (RHS above).
To recover the original claim, we set $\ket{\mathbf{y}}=\ket{\mathbf{x}}$ in order to bias towards $\ket{0\dots 0}$.

\subsubsection{Noise-Directed WS-QAOA}
Above we have shown that instead of creating WS ansatz biased towards solution $\y$ with $H_{PS}$ as the phase separator, we can implement WS ansatz biased towards $\ket{0\dots 0}$ state with $H_{PS}^{\y}$ as the phase separator; both settings being equivalent in noiseless circuits.
In the noisy setting, however, we expect the gauge-transformed ansatz to perform better for certain types of biased noise, such as amplitude damping.
Heuristically, this is because in the gauge-transformed setup, the ansatz always biases towards the $\ket{0\dots 0}$, which is the same direction as the noise bias.
In the non-gauge-transformed version with changing bias towards either $\ket{0}$ or $\ket{1}$, the noise will sometimes act against the bias introduced by the ansatz (this will happen whenever $f_i\left(x_i\right)>\frac{1}{2}$).
Indeed, in Refs.~\cite{maciejewski2024ndar,maciejewski2024multilevel,tam2025enhancingNDAR}, the authors use Noise-Directed Adaptive Remapping that relies solely on amplitude damping effects to perform iterative optimization.

Throughout the paper, we consider the WS-ansatz biased towards $\ket{0 \dots 0}$, understanding that biasing towards an arbitrary solution can be easily achieved by gauge-transforming the cost Hamiltonian.
Besides the experimentally demonstrated noise-adaptivity, we believe that the gauge-transformed picture offers practical implementation simplicity, as one always constructs a quantum circuit that biases towards the same physical state. 
Moreover, we believe that thinking in terms of gauge-transformed WS ansatz is conceptually beneficial, as one can always analyze the behavior around the $\ket{0\dots 0}$ state.
We refer to such an ansatz as a \emph{Noise-Directed (ND)} WS ansatz.

\subsection{Note on more general ND-AWS  variants}\label{app:theory:IWS-general}

In Eq.~\eqref{eq:ws_ininital_state} and Eq.~\eqref{eq:ws_mixer_one_qubit} in the main text, we considered the simplest variant of creating a biased ansatz, with a single parameter $c$ controlling the overall bias towards the candidate solution.
Naturally, more sophisticated variants are possible, including adaptive strategies. 
For example, the bias parameters could depend on the ND-AWS iteration -- as we gain confidence in the value of some variable, we could decrease the value of $c$; or we could introduce ``annealing schedules" that decrease $c$ (``lower the temperature'') with iteration number (we implemented simplified version of this strategy in experiments presented in Section~\ref{sec:exp}); or in the case of ND-AWS getting stuck, we could increase $c$ to increase the size of the search space.
Another modification could involve choosing the value of $c_i$ ($i$ labeling a qubit) based on the value of the local expected value $\left<Z_i\right>$ optimized in the previous iteration step. 
The idea is similar to iterative quantum optimization~\cite{brady2023iterative}, where we interpret a higher value of $|\left<Z_i\right>|$ as a higher ``certainty" of the solver that variable $x_i$ should be either $0$ or $1$ (this choice mainly depends on whether one is a physicist or a computer scientist). 
We expect similar modifications to improve the overall performance of the algorithm.

\subsection{Time-Block QAOA}\label{app:time_block_ansatz}

For large, dense problems, it is typically impractical to implement even a single-layer $p=1$ QAOA ansatz on hardware. 
To alleviate this difficulty, in our experiments, we implement the Time-Block (TB) QAOA ansatz (and its Warm-Started variant) introduced in Ref.~\cite{maciejewski2023design} (see also Refs.~\cite{wilkie2024qaoa,wang2025linearchainQAOA,koch2025decathlon} for similar ideas).
The idea of the Time-Block ansatz is to divide a single layer of QAOA into multiple sub-layers, each with a different phase separation operator (followed by a standard mixer) corresponding to only a \emph{subset} of total Hamiltonian interactions. 

Whenever the union of all interaction subsets in TB-QAOA constitutes the whole Hamiltonian, one can view this ansatz as an over-parametrized generalization of the QAOA -- the original ansatz is recovered by setting all phase-separator angles to the same value, and all-but-last mixer angles to $0$.

However, here we are interested in the \emph{under}-parametrized version of TB-QAOA, which intentionally does not implement all of the Hamiltonian interactions.
In a noiseless setting, this, in general, will lead to worse performance than vanilla QAOA if not enough interactions are included in the ansatz~\cite{maciejewski2023design}.
However, in the noisy, experimental setting, implementing full $p=1$ QAOA can lead to overwhelming noise effects even for problems of moderate size.
Motivated by this, in this work, we experimentally implemented the $p=1$ underparametrized Time-Block QAOA ansatz (and its Warm-Started variants).
In our $100$-qubit experiments, for $10\%$-dense Erd\H{o}s-R\'enyi graphs, we implemented $25\%$ of all interactions in the phase separator, while for $20\%$-dense ER, only $10\%$ of all interactions; and $80\%$ for RG-3.
This resulted in circuit depths summarized in Table~\ref{tab:summary_implementation}.

Consider Hamiltonian with edges set $\left\{\left(i,j\right) \in \mathcal{S}\right\}$.
Specifying division $\mathcal{S} = \cup_{R}\ \mathcal{S}_R$ defines Hamiltonian batching via
\begin{equation}
    H = \sum_{R}H_R= \sum_{R} \sum_{\left(i,j\right)\in \mathcal{S}_R} J_{i,j}Z_iZ_j \ .
\end{equation}
The Time-Block (TB) ansatz introduced in Ref.~\cite{maciejewski2023design} proposes to treat each $H_R$ as a Phase Separator (PS) operator in the QAOA layers.
Each PS layer corresponding to $H_R$ is followed by the standard QAOA mixer.
This allows for both under- and over-parametrization compared to vanilla QAOA.
Let us denote $k$-Time-Block with $k = \frac{|S_R|}{|S|}$ as the fraction of interactions implemented in a single PS layer.
Here for simplicity we assume that $k$ does not depend on index $R$ -- in practice, the last layer might be  larger than the rest to cover the whole graph. 
Since we use only $p=1$ TB-QAOA in our experiments, let us focus on this case.
We have a single-element division $\mathcal{S}_1$ and whenever $\mathcal{S}_{1}\neq\mathcal{S}$, only a fraction of the interactions is implemented.
In this case, whenever $k<1.0$, TB-QAOA is equivalent to implementing a single vanilla QAOA layer with a less dense Phase Separator Hamiltonian.
As mentioned above, in an ideal setting, this generally should lead to worse performance than vanilla QAOA. In experiments, this leads to shallower circuits -- in general, one observes an expressivity/noise robustness tradeoff when choosing $k$. 

In Ref.~\cite{maciejewski2023design}, the division $\cup_{R}\ \mathcal{S}_R$ implemented experimentally was chosen to coincide with the Linear Swap Network (LSN) structure using construction from Ref.~\cite{hirata2011}.
Specifically, each $\mathcal{S}_R$ subset was related to an integer parameter $k \in \left\{1,2,\dots,n\right\}$ SWAP layers, i.e., chains of $ZZ$ interactions followed by SWAPs.
Hamiltonians considered in Ref.~\cite{maciejewski2023design} were fully-connected, which motivated relating the Hamiltonian batching to LSN. 

In this work, we often consider much sparser Hamiltonians, and the routing is done using the SABRE~\cite{li2019sabre,zou2024lightsabre} algorithm and $0<k\leq 1.0$, as opposed to LSN. 
Therefore, in our experiments with $p=1$, we choose the first $k|S|$ largest interactions for each Hamiltonian instance for the first $4$ iterations and random subsets afterwards, implemented with the fractional batching described above.
For fixed Hamiltonian instance, the random subsets are fixed for each iteration index, independent of the gauge, meaning that all experiment variants implement the same subset of Hamiltonian interactions at given NDAR iteration.

\subsection{Optimization strategy}

For all simulations and experiments presented in the text, we effectively reduced $p=1$ angles' optimization to single-parameter optimization, because for fixed phase-separator angle $\gamma$, it is cheap to scan over mixer angle $\beta$ (as discussed in more detail in Section~\ref{app:numerics:large_scale_p1} below).
To optimize $\gamma$, we implemented \texttt{COBYQA} optimizer~\cite{Ragonneau2023COBYQA} with basinhopping~\cite{wales1997basinhopping}, implementation from \texttt{scipy}~\cite{scipy}.
The non-default hyperparameters included $\texttt{initial\_tr\_radius}=0.01$ and $\texttt{final\_tr\_radius}=10^{-5}$ for \texttt{COBYQA}; as well as $\texttt{niter}=5$ and $\texttt{T}=2.0$ for \texttt{basinhopping}.
Each optimization was started from the $\gamma_0=0.0001$ point, and was capped at $200$ cost function evaluations in total (i.e., over all local optimizer runs).

\section{Supporting numerical results}\label{app:more_numerics}

In this appendix, we present further numerical studies of $p=1$ ND-AWS with $H_\mathrm{PS}=H_C$, complementing the noise-adaptivity study of Section~\ref{sec:numerics}.
Appendix~\ref{sec:numerics:mps} presents $\noq=100$ qubit approximate simulations via MPS.
Appendix~\ref{app:numerics:large_scale_p1} presents a study of $p=1$ ND-AWS with $500$-qubit problems based on the efficient computation of expected values, a regime where direct sampling becomes computationally challenging.

\subsection{Large-scale Approximate Noiseless Simulations With Matrix Product States}\label{sec:numerics:mps}

To show how quantum effects impact result quality we now present noiseless simulations with samples drawn from an MPS approximation, done with \texttt{qiskit\_aer}~\cite{qiskit2024}, at different bond dimensions of $p=1$ ND-AWS.
For efficiency, we optimize $\beta$ and $\gamma$ with the fast and exact $p=1$ expected-values simulator. 
We consider the same $10$ random instances as in the experiments from Section~\ref{sec:exp} with an identical schedule for the bias parameter $c$. 
However, the phase separator in the simulations is equal to the cost Hamiltonian, unlike in the experiments, where we truncated the implemented interactions.

\begin{figure*}[t!]
\centering
\includegraphics[width=\textwidth]{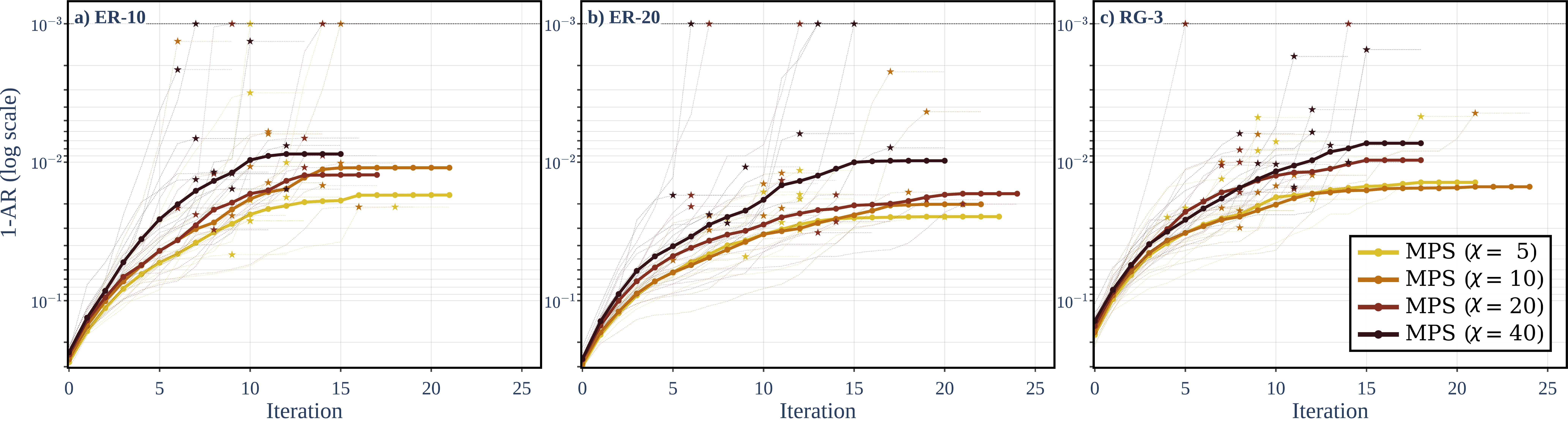}
\caption{\label{fig:sim:mps_ndar}
Distance to the optimal solution, $1-AR$, of the best-found sample (logarithmic scale; inverted axis as in Fig.~\ref{fig:sim:small_scale_performance}) at each iteration of the $p=1$ ND-AWS for $\noq=100$-qubit random Erd\H{o}s-R\'enyi graphs ER-10 \textbf{(a)} and ER-20 \textbf{(b)}, and random 3-regular graphs RG-3 \textbf{(c)}, with samples drawn from an MPS approximation.
Colors correspond to different bond dimensions $\chi$; the legend in \textbf{(c)} applies to all panels.
Each thin dotted line is a different random Hamiltonian instance, chosen as the best out of $3$ independent runs, with its best point indicated by a star symbol; optimal solutions ($AR=1$) are clipped to $10^{-3}$ (dotted horizontal line).
Thick solid lines with circle markers show the mean over all instances.
}
\end{figure*}

Each simulation is repeated for $3$ different seeds, and we show the best out of those $3$ runs. 
Each simulation is done for bond dimensions $\chi \in \left\{5, 10, 20, 40\right\}$.
On average, increasing bond dimension leads to better quality of results, smaller spread, and faster convergence; see Fig.~\ref{fig:sim:mps_ndar}. 
This shows that allowing more entanglement in the circuit increases quality.

\subsection{Large-scale Exact Simulations With $p=1$ Expected Values}\label{app:numerics:large_scale_p1}

Here, we present exact larger-scale noiseless simulations based on computing expectation values. These quantities reasonably characterize the expected performance of our approach in the noiseless setting; in practice, sampling should enable even better performance, as suggested by the MPS-based approximate simulations discussed in Appendix~\ref{sec:numerics:mps}.

\subsubsection{Simulation algorithm}

To calculate expected values of two-local Hamiltonians for $p=1$ Warm Start QAOA, we use a slightly modified version of Algorithm~1 from Appendix F of Ref.~\cite{egger2021warm}.
We assume that bias parameters are identical for every qubit, so that all mixer operators are the same. 
In summary, the algorithm for that case goes as follows: 
\begin{enumerate}
    \item For fixed $\gamma$, efficiently calculate all 2-body reduced density matrices (RDMs) $\rho_{i,j}$.
    \item For fixed $\beta$, apply the product mixer $U_M\left(\beta\right)^{\otimes 2}$ to all RDMS.
    \item Calculate overlaps $\mathrm{Tr}\left(ZZ\ U_M\left(\beta\right)^{\otimes 2}\rho_{i,j}U_M\left(\beta\right)^{\otimes 2}\right)$.
\end{enumerate}
The most computationally-intensive step 1 above is straightforward for WS-QAOA circuits, as shown in Ref.~\cite{egger2021warm}.
Note that the procedure can also be implemented for Hamiltonians with local fields, using 1-body RDMs instead (at quadratically lower complexity in general). 

In our implementation, we make a simple modification to adjust the algorithm to take into account the fact that we are interested in optimizing the cost function $\mathrm{Tr}\left(\rho H\right)$, and the values of particular correlators are not needed.
In particular, we note that
\begin{equation}
    \begin{split}
        \mathrm{Tr}\left(\rho H\right) &= \sum_{i<j} J_{i,j} \mathrm{Tr}\left(ZZ\ U_M\left(\beta\right)^{\otimes 2}\rho_{i,j}U_M\left(-\beta\right)^{\otimes 2}\right) \\ &=  \mathrm{Tr}\left(U_M\left(-\beta\right)^{\otimes 2}ZZ\ U_M\left(\beta\right)^{\otimes 2}\sum_{i<j} J_{i,j} \rho_{i,j}\right) \\
        &=\mathrm{Tr}\left(\tilde{ZZ}\ \bar{\rho}\right) \ ,
    \end{split}
\end{equation}
where in the second line we used facts that trace is a cyclic and linear functional, while in the last we defined the Heisenberg-picture image of the $ZZ$ operator under the two-body mixer 
\begin{equation}
    \tilde{ZZ} \coloneqq U_M\left(-\beta\right)^{\otimes 2}ZZ\ U_M\left(\beta\right)^{\otimes 2} \ ,
\end{equation}
and a ``weighted average" 2-body RDM
\begin{equation}
    \bar{\rho} \coloneqq \sum_{i<j} J_{i,j} \rho_{i,j} \ .
\end{equation}
For fixed $\gamma$, the weighted average $\bar{\rho}$ is computed once. 
Then, evaluating overlaps of $\tilde{ZZ}$ with that operator for any $\beta$ involves operations on $4\times 4$ matrices.
This can be done very efficiently, allowing for finding numerically optimal $\beta$ for any fixed $\gamma$.
In practice, in our optimization, for each $\gamma$, we calculated expected values for $\geq 10^6$ distinct values of $\beta$ (giving grid-search resolution $\leq \frac{\pi}{10^6}$), at little additional computational cost.
We note that this approach is conceptually similar to Ref.~\cite{vijendran2025near}, which investigated finding analytically-optimal $\beta$ for each $\gamma$ in the vanilla (no Warm-Start) $p=1$ setting.

In the above, we assumed that the mixer does not depend on the qubits, meaning a fixed bias parameter $c$ -- a simplification possible mainly due to use of gauge-transformations that move the bias change onto the phase separator (recall Appendix~\ref{app:qaoa_symmetries}).
In the case of qubit-dependent bias, the algorithm would have to be trivially modified to include qubit-dependent $\tilde{ZZ}$ operators.
Our $p=1$ Noise-Directed WS-QAOA simulator based on the \texttt{numba.cuda} package~\cite{lam2015numba} will be available in an open-source repository \texttt{quapopt}~\cite{quapopt_repo}.

\subsubsection{Results}

We study how the optimized \emph{mean} approximation ratio of the $p=1$ ND-WS-QAOA ansatz depends on the bias parameter $c$ and on the quality of the candidate solution (quantified by the AR of the $\ket{0\dots 0}$ state, to which the candidate is logically equivalent after the gauge transformation).
The simulations are performed for $10$ random instances of $500$-qubit ER-10 Hamiltonians; for each candidate solution, the variational angles are optimized and the best result over $8$ bias-parameter values (including $c=0$ and the no-bias vanilla value $c=0.5$) is selected.
Two baselines are relevant: the AR of standard ND-TB-QAOA without bias ($c=0.5$), and the AR of the candidate solution itself.
Note that if the optimized \emph{mean} AR exceeds the AR of the candidate solution, the underlying distribution necessarily has non-vanishing support on better solutions; the $p=1$ expected-value simulation can therefore positively verify whether WS-QAOA is capable of sampling improved solutions.
On the other hand, a distribution whose mean AR lies only slightly below that of the candidate can still yield improved solutions upon sampling; the likelihood of this depends on higher moments of the distribution, which are much harder to simulate classically.

\begin{wrapfigure}{r}{0.46\textwidth}
\centering
\includegraphics[width=0.45\textwidth]{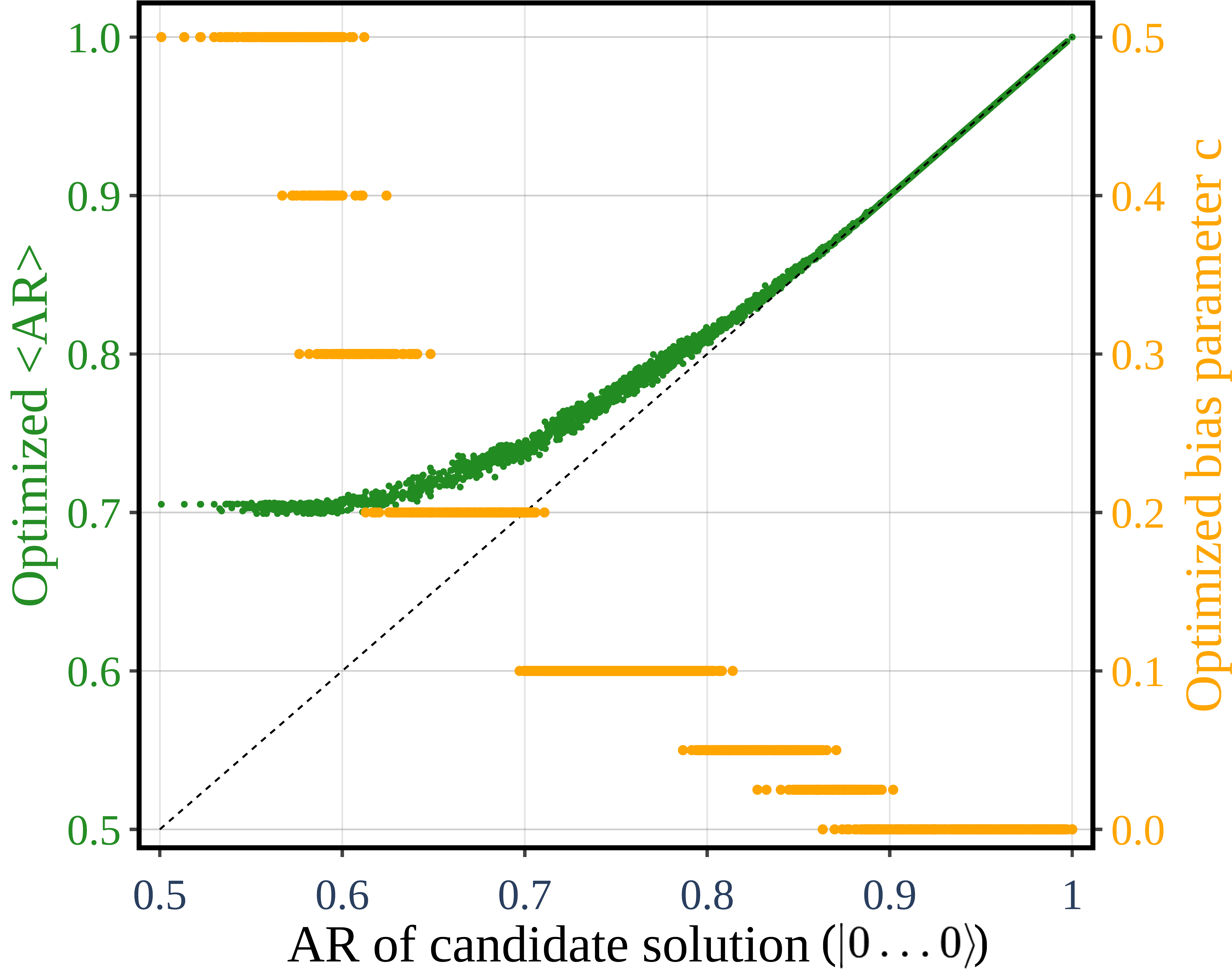}
\caption{\label{fig:sim:large_scale_p1_optimized_bias_and_ar}
Optimized mean Approximation Ratio of $p=1$ Noise-Directed Warm-Start QAOA (left Y-axis; green points) and the corresponding optimized Bias Parameter (right Y-axis; orange points), as a function of the candidate solution quality (X-axis), aggregated over $10$ random instances of $500$-qubit ER-10 Hamiltonians.
The dashed line corresponds to the $y=x$ line w.r.t.\ the optimized $\left<AR\right>$ axis: points above it indicate that the optimized state improves, on average, upon the candidate solution.
}
\end{wrapfigure}
The results (aggregate over all $10$ instances) are presented in Fig.~\ref{fig:sim:large_scale_p1_optimized_bias_and_ar}.
We plot AR of $\ket{0\dots0}$ against mean optimized AR (green), and indicate which bias $c$ gave the best performance with the second Y axis (orange points); i.e., each green datapoint corresponds to the \emph{best} mean AR, where the maximum is taken over $8$ values of the bias parameter $c$.
We make the following observations.
First, the optimized warm-start ansatz outperforms the no-bias baseline whenever the candidate solution is good enough, with the transition occurring at a candidate AR of $\approx 0.6$--$0.645$ across all considered instances; for lower-quality candidates, the optimized bias approaches $c=0.5$ and warm-starting offers no advantage.
Second, the optimized mean AR exceeds the AR of the candidate solution itself (points above the dotted diagonal in Fig.~\ref{fig:sim:large_scale_p1_optimized_bias_and_ar}) for candidate ARs up to a plateau at $\approx 0.85$--$0.9$, indicating that noiseless sampling from the optimized state would likely improve upon the input solution in this regime.
Those results suggest that our warm-starting remains effective at scales ($n=500$) well beyond our experiments, even when using only depth-1 WS-QAOA.
Finally, the best-found bias (orange staircase in Fig.~\ref{fig:sim:large_scale_p1_optimized_bias_and_ar}) decreases monotonically as the candidate quality grows.
This is a strong indication in favor of an ``annealing'' strategy for the bias parameter $c$, where, with increasing quality of the best-guess solution, we decrease $c$ to perform a more localized exploration of the search space.
In the QPU experiments of Section~\ref{sec:exp}, we indeed use a simple version of this strategy.
Taken together, these observations indicate that single-step improvements can be chained: each accepted solution lands in a regime where a smaller bias still offers further improvement, so that repeated iterations with an annealed bias schedule progressively walk towards higher-quality solutions -- the mechanism realized experimentally in Section~\ref{sec:exp}.
Moreover, in the underlying scans over the eight bias values, we observe that in the intermediate-quality regime the optimized mean AR exceeds both baselines simultaneously over a contiguous range of bias values, $0.025 \lesssim c \lesssim 0.2$, which covers the values used in our experimental schedule; the advantage therefore does not seem to require precise fine-tuning of $c$.

\section{ND-AWS experiments}

\subsection{QPU implementation}\label{app:hardware_implementation}

All experiments reported in the main text were performed on $100$-qubit subsystems of \texttt{ibm\_boston} device. 
The native gate set of the device consists of SX and (virtual) RZ single-qubit gates and the CZ entangling gate. 

Before running any circuit on the QPU, we queried the backend for the most recent calibration data on fidelities of the native gates, as well as readout errors. 
We calculated the mean and standard deviation of each figure of merit.
Then, we excluded qubits and edges further away than $2\sigma$ from the mean (in the direction of worse performance) from the coupling map passed to the transpiler. 

The transpilation was performed with \texttt{qiskit}'s pass manager with \texttt{optimization\_level}=3 and \texttt{routing\_method}=`sabre' parameters with seed set to $42$.

\section{Additional experiments without greedy post-processing}\label{app:hdls_runs}

\begin{table*}[]
\centering
\resizebox{\textwidth}{!}{%
\begin{tabular}{cclcccclcccclc}
\cline{1-2} \cline{4-7} \cline{9-12} \cline{14-14}
\multicolumn{1}{|l|}{\multirow{2}{*}{\textbf{Hamiltonian}}} & \multicolumn{1}{c|}{\multirow{2}{*}{\textbf{Ansatz}}} & \multicolumn{1}{l|}{} & \multicolumn{4}{c|}{\textbf{Iterations at convergence}}                                                                                                                         & \multicolumn{1}{l|}{} & \multicolumn{4}{c|}{\textbf{Best AR at convergence}}                                                                                                                               & \multicolumn{1}{l|}{} & \multicolumn{1}{c|}{\textbf{AR at iteration 10}} \\ \cline{4-7} \cline{9-12} \cline{14-14} 
\multicolumn{1}{|l|}{}                                      & \multicolumn{1}{c|}{}                                 & \multicolumn{1}{l|}{} & \multicolumn{1}{l|}{\textbf{min}} & \multicolumn{1}{l|}{\textbf{max}} & \multicolumn{1}{c|}{\textbf{median}} & \multicolumn{1}{c|}{\textbf{mean ($\pm$ SD)}}     & \multicolumn{1}{l|}{} & \multicolumn{1}{l|}{\textbf{min}} & \multicolumn{1}{l|}{\textbf{max}} & \multicolumn{1}{c|}{\textbf{median}} & \multicolumn{1}{c|}{\textbf{mean ($\pm$ SD)}}        & \multicolumn{1}{l|}{} & \multicolumn{1}{c|}{\textbf{mean ($\pm$ SD)}} \\ \cline{1-2} \cline{4-7} \cline{9-12} \cline{14-14} 
\multicolumn{1}{l}{}                                        & \multicolumn{1}{l}{}                                  &                       & \multicolumn{1}{l}{}              & \multicolumn{1}{l}{}              & \multicolumn{1}{l}{}                 & \multicolumn{1}{l}{}                              &                       & \multicolumn{1}{l}{}              & \multicolumn{1}{l}{}              & \multicolumn{1}{l}{}                 & \multicolumn{1}{l}{}                                 & & \\ \cline{1-2} \cline{4-7} \cline{9-12} \cline{14-14} 
\multicolumn{1}{|c|}{\multirow{3}{*}{ER-10}}                & \multicolumn{1}{c|}{Standard IWS}                         & \multicolumn{1}{l|}{} & \multicolumn{1}{c|}{$12$}           & \multicolumn{1}{c|}{$21$}           & \multicolumn{1}{c|}{$16$}            & \multicolumn{1}{c|}{$16.1\ \left(\pm 2.7\right)$} & \multicolumn{1}{c|}{} & \multicolumn{1}{c|}{$0.893$}         & \multicolumn{1}{c|}{$0.979$}         & \multicolumn{1}{c|}{$0.932$}          & \multicolumn{1}{c|}{$0.934\ \left(\pm 0.028\right)$}   & \multicolumn{1}{l|}{} & \multicolumn{1}{c|}{$0.913\ \left(\pm 0.04\right)$} \\ \cline{2-2} \cline{4-7} \cline{9-12} \cline{14-14} 
\multicolumn{1}{|c|}{}                                      & \multicolumn{1}{c|}{\cellcolor{black!10}ND-AWS}                   & \multicolumn{1}{l|}{\cellcolor{black!10}} & \multicolumn{1}{c|}{\cellcolor{black!10}$13$}           & \multicolumn{1}{c|}{\cellcolor{black!10}$32$}           & \multicolumn{1}{c|}{\cellcolor{black!10}$16.5$}          & \multicolumn{1}{c|}{\cellcolor{black!10}$19.5\ \left(\pm 6.2\right)$} & \multicolumn{1}{c|}{\cellcolor{black!10}} & \multicolumn{1}{c|}{\cellcolor{black!10}$0.947$}         & \multicolumn{1}{c|}{\cellcolor{black!10}$1.0$}          & \multicolumn{1}{c|}{\cellcolor{black!10}$0.970$}          & \multicolumn{1}{c|}{\cellcolor{black!10}$0.976\ \left(\pm 0.019\right)$}   & \multicolumn{1}{l|}{\cellcolor{black!10}} & \multicolumn{1}{c|}{\cellcolor{black!10}$0.954\ \left(\pm 0.016\right)$} \\ \cline{2-2} \cline{4-7} \cline{9-12} \cline{14-14} 
\multicolumn{1}{|c|}{}                                      & \multicolumn{1}{c|}{MPS}                              & \multicolumn{1}{l|}{} & \multicolumn{1}{c|}{$9$}            & \multicolumn{1}{c|}{$15$}           & \multicolumn{1}{c|}{$11.5$}            & \multicolumn{1}{c|}{$11.8\ \left(\pm 1.9\right)$}   & \multicolumn{1}{l|}{} & \multicolumn{1}{c|}{$0.984$}         & \multicolumn{1}{c|}{$1.0$}          & \multicolumn{1}{c|}{$0.990$}          & \multicolumn{1}{c|}{$0.991\ \left(\pm 0.006\right)$}   & \multicolumn{1}{l|}{} & \multicolumn{1}{c|}{$0.990\ \left(\pm 0.007\right)$} \\ \cline{1-2} \cline{4-7} \cline{9-12} \cline{14-14} 
\multicolumn{1}{l}{}                                        & \multicolumn{1}{l}{}                                  &                       & \multicolumn{1}{l}{}              & \multicolumn{1}{l}{}              & \multicolumn{1}{l}{}                 & \multicolumn{1}{l}{}                              &                       & \multicolumn{1}{l}{}              & \multicolumn{1}{l}{}              & \multicolumn{1}{l}{}                 & \multicolumn{1}{l}{}                                 & & \\ \cline{1-2} \cline{4-7} \cline{9-12} \cline{14-14} 
\multicolumn{1}{|c|}{\multirow{3}{*}{ER-20}}                & \multicolumn{1}{c|}{Standard IWS}                         & \multicolumn{1}{l|}{} & \multicolumn{1}{c|}{$12$}           & \multicolumn{1}{c|}{$29$}           & \multicolumn{1}{c|}{$25$}            & \multicolumn{1}{c|}{$22.9\ \left(\pm 5.9\right)$} & \multicolumn{1}{c|}{} & \multicolumn{1}{c|}{$0.944$}         & \multicolumn{1}{c|}{$0.996$}        & \multicolumn{1}{c|}{$0.970$}          & \multicolumn{1}{c|}{$0.970\ \left(\pm 0.017\right)$}   & \multicolumn{1}{l|}{} & \multicolumn{1}{c|}{$0.947\ \left(\pm 0.027\right)$} \\ \cline{2-2} \cline{4-7} \cline{9-12} \cline{14-14} 
\multicolumn{1}{|c|}{}                                      & \multicolumn{1}{c|}{\cellcolor{black!10}ND-AWS}                   & \multicolumn{1}{l|}{\cellcolor{black!10}} & \multicolumn{1}{c|}{\cellcolor{black!10}$9$}            & \multicolumn{1}{c|}{\cellcolor{black!10}$25$}           & \multicolumn{1}{c|}{\cellcolor{black!10}$14$}            & \multicolumn{1}{c|}{\cellcolor{black!10}$15.5\ \left(\pm 4.4\right)$} & \multicolumn{1}{c|}{\cellcolor{black!10}} & \multicolumn{1}{c|}{\cellcolor{black!10}$0.949$}         & \multicolumn{1}{c|}{\cellcolor{black!10}$0.997$}        & \multicolumn{1}{c|}{\cellcolor{black!10}$0.975$}          & \multicolumn{1}{c|}{\cellcolor{black!10}$0.975\ \left(\pm 0.015\right)$}   & \multicolumn{1}{l|}{\cellcolor{black!10}} & \multicolumn{1}{c|}{\cellcolor{black!10}$0.966\ \left(\pm 0.015\right)$} \\ \cline{2-2} \cline{4-7} \cline{9-12} \cline{14-14} 
\multicolumn{1}{|c|}{}                                      & \multicolumn{1}{c|}{MPS}                              & \multicolumn{1}{l|}{} & \multicolumn{1}{c|}{$8$}            & \multicolumn{1}{c|}{$20$}           & \multicolumn{1}{c|}{$13.5$}          & \multicolumn{1}{c|}{$13.7\ \left(\pm 4.0\right)$} & \multicolumn{1}{l|}{} & \multicolumn{1}{c|}{$0.973$}         & \multicolumn{1}{c|}{$1.0$}          & \multicolumn{1}{c|}{$0.993$}          & \multicolumn{1}{c|}{$0.991\ \left(\pm 0.010\right)$}   & \multicolumn{1}{l|}{} & \multicolumn{1}{c|}{$0.981\ \left(\pm 0.009\right)$} \\ \cline{1-2} \cline{4-7} \cline{9-12} \cline{14-14} 
\multicolumn{1}{l}{}                                        & \multicolumn{1}{l}{}                                  &                       & \multicolumn{1}{l}{}              & \multicolumn{1}{l}{}              & \multicolumn{1}{l}{}                 & \multicolumn{1}{l}{}                              &                       & \multicolumn{1}{l}{}              & \multicolumn{1}{l}{}              & \multicolumn{1}{l}{}                 & \multicolumn{1}{l}{}                                 & & \\ \cline{1-2} \cline{4-7} \cline{9-12} \cline{14-14} 
\multicolumn{1}{|c|}{\multirow{3}{*}{RG-3}}                 & \multicolumn{1}{c|}{Standard IWS}                         & \multicolumn{1}{l|}{} & \multicolumn{1}{c|}{$10$}           & \multicolumn{1}{c|}{$27$}           & \multicolumn{1}{c|}{$17$}            & \multicolumn{1}{c|}{$16.5\ \left(\pm 5.0\right)$} & \multicolumn{1}{c|}{} & \multicolumn{1}{c|}{$0.962$}         & \multicolumn{1}{c|}{$0.9985$}        & \multicolumn{1}{c|}{$0.979$}          & \multicolumn{1}{c|}{$0.979\ \left(\pm 0.011\right)$}   & \multicolumn{1}{l|}{} & \multicolumn{1}{c|}{$0.972\ \left(\pm 0.015\right)$} \\ \cline{2-2} \cline{4-7} \cline{9-12} \cline{14-14} 
\multicolumn{1}{|c|}{}                                      & \multicolumn{1}{c|}{\cellcolor{black!10}ND-AWS}                   & \multicolumn{1}{l|}{\cellcolor{black!10}} & \multicolumn{1}{c|}{\cellcolor{black!10}$9$}            & \multicolumn{1}{c|}{\cellcolor{black!10}$21$}           & \multicolumn{1}{c|}{\cellcolor{black!10}$15$}            & \multicolumn{1}{c|}{\cellcolor{black!10}$14.6\ \left(\pm 3.6\right)$} & \multicolumn{1}{c|}{\cellcolor{black!10}} & \multicolumn{1}{c|}{\cellcolor{black!10}$0.969$}         & \multicolumn{1}{c|}{\cellcolor{black!10}$0.997$}        & \multicolumn{1}{c|}{\cellcolor{black!10}$0.981$}          & \multicolumn{1}{c|}{\cellcolor{black!10}$0.981\ \left(\pm 0.008\right)$}   & \multicolumn{1}{l|}{\cellcolor{black!10}} & \multicolumn{1}{c|}{\cellcolor{black!10}$0.975\ \left(\pm 0.007\right)$} \\ \cline{2-2} \cline{4-7} \cline{9-12} \cline{14-14} 
\multicolumn{1}{|c|}{}                                      & \multicolumn{1}{c|}{MPS}                              & \multicolumn{1}{l|}{} & \multicolumn{1}{c|}{$11$}           & \multicolumn{1}{c|}{$18$}           & \multicolumn{1}{c|}{$14.5$}          & \multicolumn{1}{c|}{$14.5\ \left(\pm 2.1\right)$} & \multicolumn{1}{l|}{} & \multicolumn{1}{c|}{$0.985$}        & \multicolumn{1}{c|}{$0.998$}        & \multicolumn{1}{c|}{$0.993$}          & \multicolumn{1}{c|}{$0.993\ \left(\pm 0.004\right)$} & \multicolumn{1}{l|}{} & \multicolumn{1}{c|}{$0.988\ \left(\pm 0.005\right)$} \\ \cline{1-2} \cline{4-7} \cline{9-12} \cline{14-14} 
\end{tabular}}
\caption{\label{tab:app:ndar_performance_hdqs}
Table summarizing the data from Fig.~\ref{fig:app:ndar_hdqs} for the iterative implementation of Standard and Noise-Directed WS ansätze (QPU), as well as MPS simulations; without HDQS. 
For a fixed Hamiltonian class and ansatz, the indicated functions are applied over random Hamiltonian instances.
The last column reports the mean ($\pm$ SD over instances) of the best-found AR at iteration $10$; for runs that converged before iteration $10$, the final value is carried over.}
\end{table*}

In Figure~\ref{fig:exp:ndar_performance} discussed in Section~\ref{sec:exp}, we presented results augmented with Hamming Distance Quadratic Search (HDQS). 
For clarity, we write down the HDQS algorithm explicitly. At each iteration:
\begin{enumerate}
    \item Take all solutions from QPU $\left\{\mathbf{x}\right\}$.
    \item Generate $\noq$ bitstrings by flipping each bit $x_i$ of each $\mathbf{x}$ and $|E|$ bitstrings by flipping each pair $x_ix_j$ for $\left(i,j\right)\in E$, i.e., edges present in the Hamiltonian.
    \item Choose the bitstring that gives the lowest energy as your new candidate solution.
\end{enumerate}

\begin{figure}[t!]
\centering
\includegraphics[width=\columnwidth]{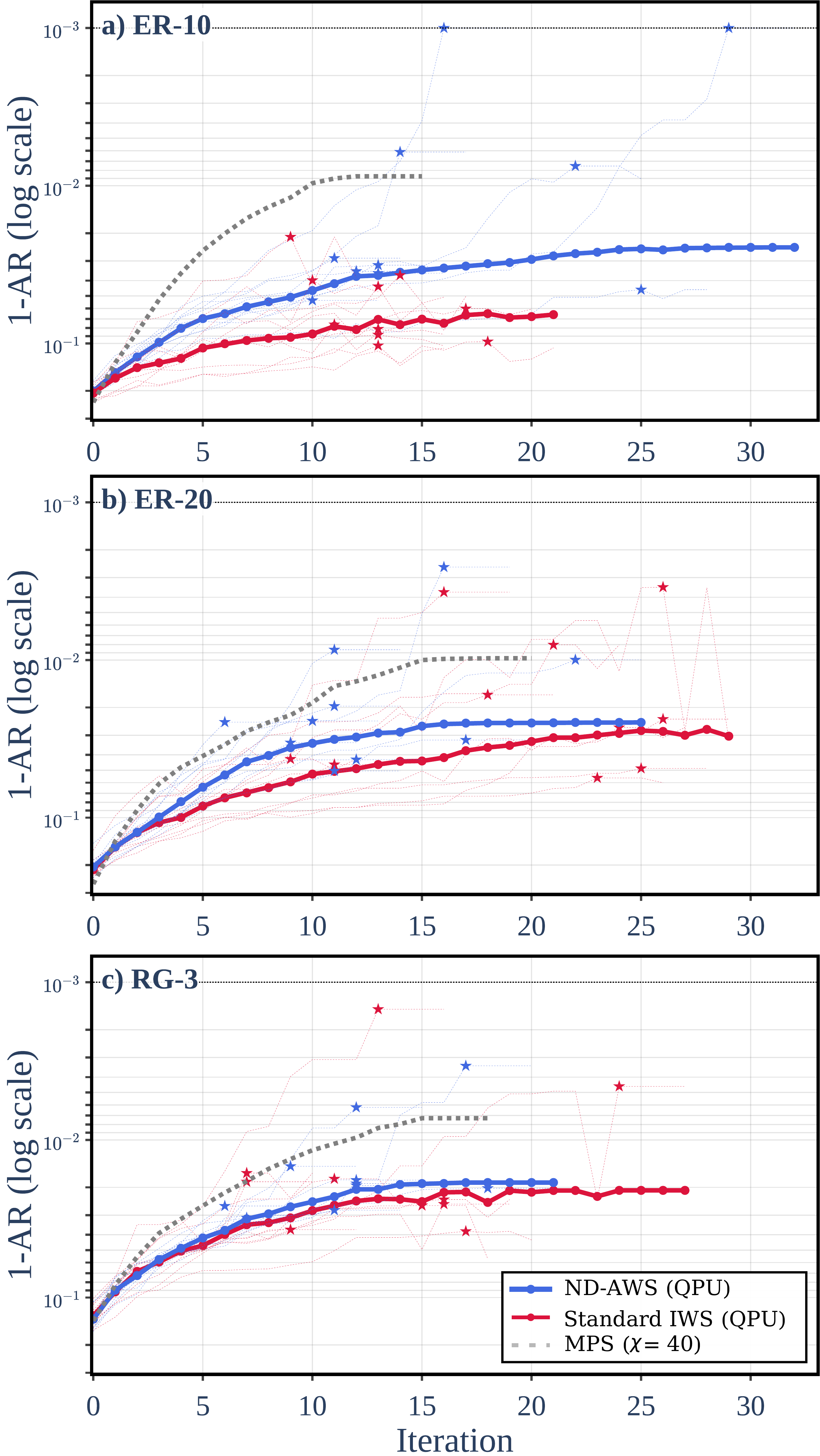}
\caption{\label{fig:app:ndar_hdqs}
Distance to the optimal solution, $1-AR$ (logarithmic scale; inverted axis).
The implemented experiments and data conventions are the same as in Fig.~\ref{fig:exp:ndar_performance} in the main text, but here we do not use additional classical post-processing (HDQS).
}
\end{figure}

\begin{figure}[h!]
\centering
\includegraphics[width=\columnwidth]{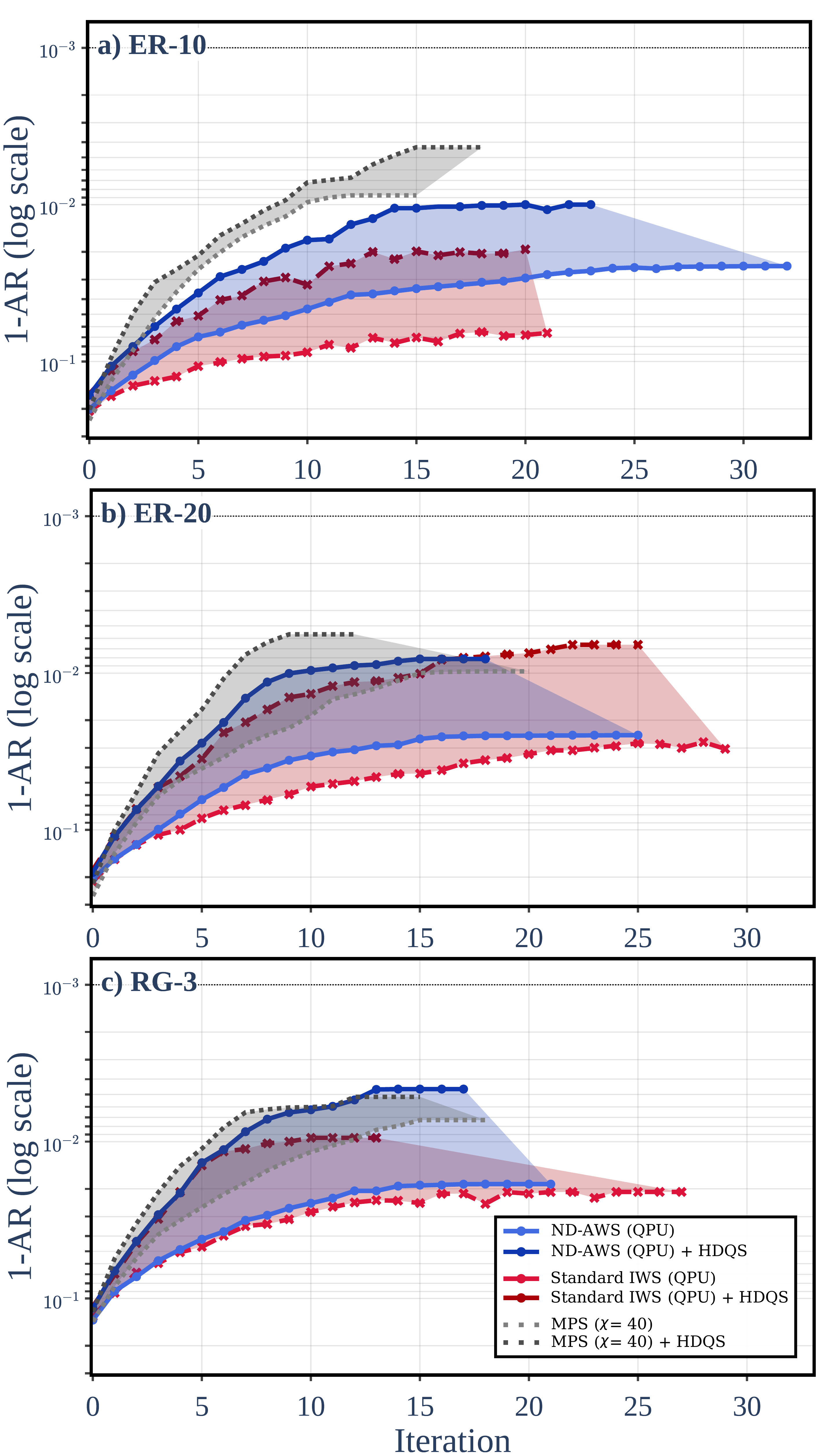}
\caption{\label{fig:app:ndar_comparison_hdqs}
Distance to the optimal solution, $1-AR$ (logarithmic scale; inverted axis).
Comparison between no-post-processing runs from Fig.~\ref{fig:app:ndar_hdqs} and HDQS-augemented runs from Fig.~\ref{fig:exp:ndar_performance}.
For clarity, only the mean values are shown.
}
\end{figure}

The results of Standard IWS and ND-AWS, as well as supporting noiseless $\chi=40$ MPS simulations without HDQS, are presented in Fig.~\ref{fig:app:ndar_hdqs}, with accompanying Fig.~\ref{fig:app:ndar_comparison_hdqs} that compares mean values obtained in no-post-processing runs (Fig.~\ref{fig:app:ndar_hdqs}) and HDQS runs from Fig.~\ref{fig:exp:ndar_performance}.
Statistics are summarizd in Table~\ref{tab:app:ndar_performance_hdqs}.
We observe that additional classical post-processing generally improves the quality of the solutions, and oftentimes the convergence time of the algorithm.

\clearpage

\twocolumngrid


\clearpage

\end{document}